\documentclass[fleqn,usenatbib]{mnras}

\usepackage[T1]{fontenc}
\usepackage{ae,aecompl}
\usepackage{gensymb}

\usepackage{graphicx}	
\usepackage{amsmath}	
\usepackage{amssymb}	

\newcommand{\photu}{photon units}
\newcommand{\mjy}{MJy sr$^{-1}$}
\newcommand{\galex}{{\it GALEX}}

\title[Components of the Diffuse Ultraviolet Radiation]{Components of the Diffuse Ultraviolet Radiation at High Latitudes}

\author[Akshaya et al.]{M. S. Akshaya,$^{1}$\thanks{E-mail: akshaya.subbanna@gmail.com}
Jayant Murthy,$^{2}$\thanks{E-mail: jmurthy@yahoo.com}
S. Ravichandran,$^{1}$\thanks{E-mail: ravichandran.s@christuniversity.in}
R. C. Henry$^{3}$\thanks{E-mail: henry@jhu.edu}
\newauthor
and James Overduin$^{4}$\thanks{E-mail: joverduin@towson.edu}
\\
$^{1}$Department of Physics and Electronics, CHRIST (Deemed to be University), Bengaluru 560 029, India\\
$^{2}$Indian Institute of Astrophysics, Bengaluru 560 034, India\\
$^{3}$Henry A. Rowland Department of Physics and Astronomy, The Johns Hopkins University, Baltimore, MD 21218, USA\\
$^{4}$Department of Physics, Astronomy and Geosciences, Towson University, Towson, MD 21252, USA\\
}

\date{\textit{Accepted for publication in MNRAS}}

\pubyear{2019}

\begin{document}
\label{firstpage}
\pagerange{\pageref{firstpage}--\pageref{lastpage}}
\maketitle

\begin{abstract}
We have used data from the {\it Galaxy Evolution Explorer} to study the different components of the diffuse ultraviolet background in the region between the Galactic latitudes 70\degree\ -- 80\degree. We find an offset at zero dust column density (E(B - V) = 0) of $240 \pm 18$ \photu\ in the FUV (1539 \AA) and $394 \pm 37$ \photu\ in the NUV (2316 \AA). This is approximately half of the total observed radiation with the remainder divided between an extragalactic component of $114 \pm 18$ \photu\ in the FUV and $194 \pm 37$ \photu\ in the NUV and starlight scattered by Galactic dust at high latitudes. The optical constants of the dust grains were found to be a=0.4$\pm$0.1 and g=0.8$\pm$0.1 (FUV) and a=0.4$\pm$0.1 and g=0.5$\pm$0.1 (NUV). We cannot differentiate between a Galactic or extragalactic origin for the zero-offset but can affirm that it is not from any known source.
\end{abstract}

\begin{keywords}
dust, extinction - ISM: clouds - diffuse radiation - ultraviolet: ISM
\end{keywords}



\section{Introduction}

The diffuse ultraviolet radiation field at the Galactic poles has been assumed to be a combination of diffuse Galactic light from the scattering of starlight by high latitude dust and extragalactic radiation but with the relative contributions unknown \citep{Bowyer1991,Henry1991}. Later observations and a better understanding of the known Galactic and extragalactic components found that they were not sufficient to account for the observed light and that there was a leftover component of 200 -- 300 \photu\footnote{photons s$^{-1}$ cm$^{-2}$ sr$^{-1}$ \AA$^{-1}$} \citep{Onaka1991, Hamden2013, Henry2015, Boissier2015}.

\citet{Akshaya2018} used observations from the \textit{Galaxy Evolution Explorer} (\textit{GALEX}) in the far ultraviolet (FUV: 1344 -- 1786 {\AA}) and the near ultraviolet (NUV: 1771 -- 2831 {\AA}) to explore the contributions of the different components at both Galactic poles ($|b| \geq 80\degree$) in conjunction with a model for the dust scattered light \citep{Murthy2016}. They found that there was an offset at zero E(B-V) of about 150 \photu\ in the FUV and 350 \photu\ in the NUV over and beyond the known extragalactic contributors. The dust scattered light was never more than about 30\% of the total signal, even at high column densities. Molecular hydrogen fluorescence kicked in for column densities of greater than log N(H) = 20.2 (where N(H) is the column density in cm$^{-2}$), significantly lower than the canonical values of log N(H)$>$20.6 \citep{Savage1977,Franco1986,Reach1994}.

\begin{figure*}
	\includegraphics[scale=0.38]{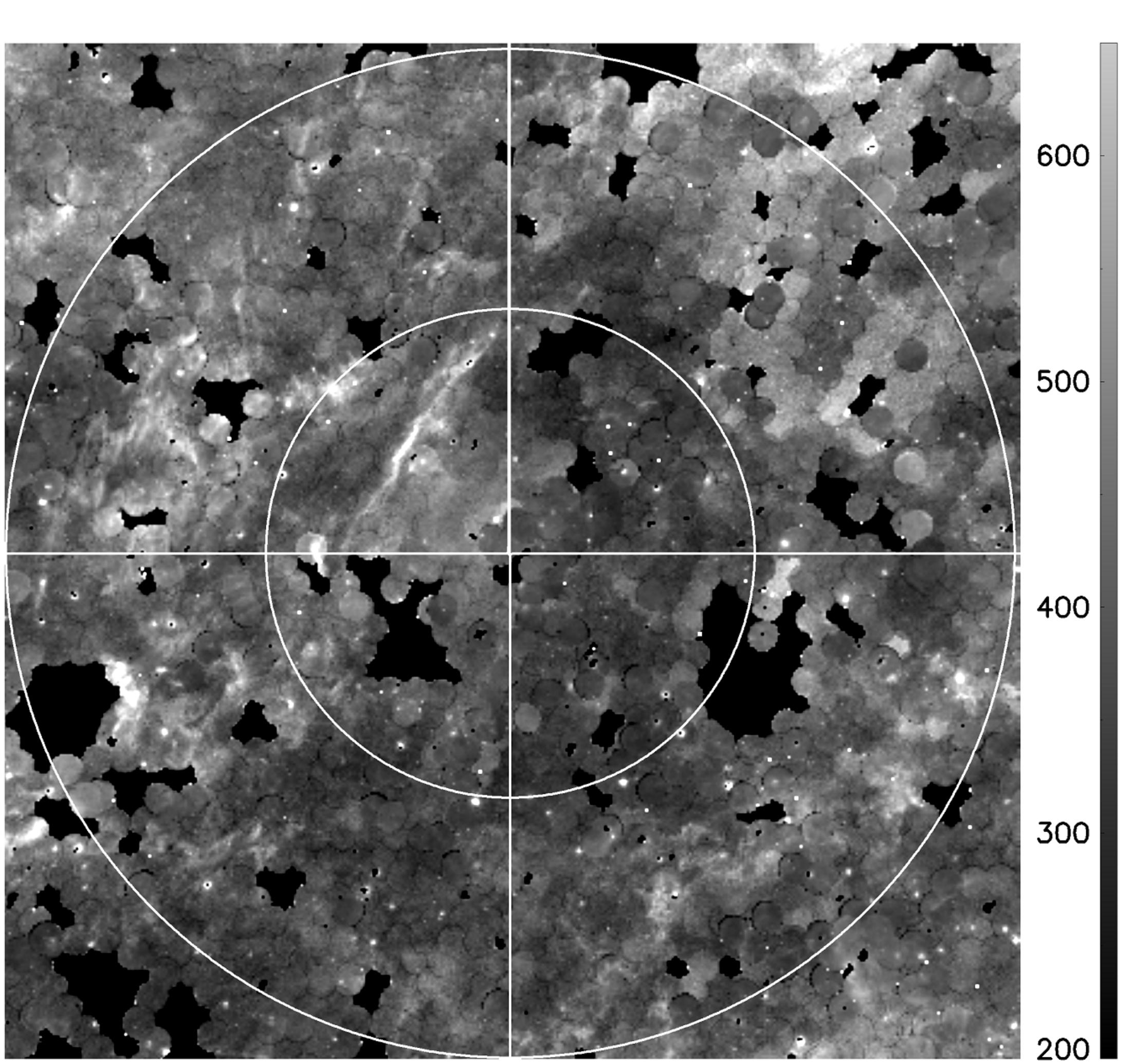}
    \includegraphics[scale=0.38]{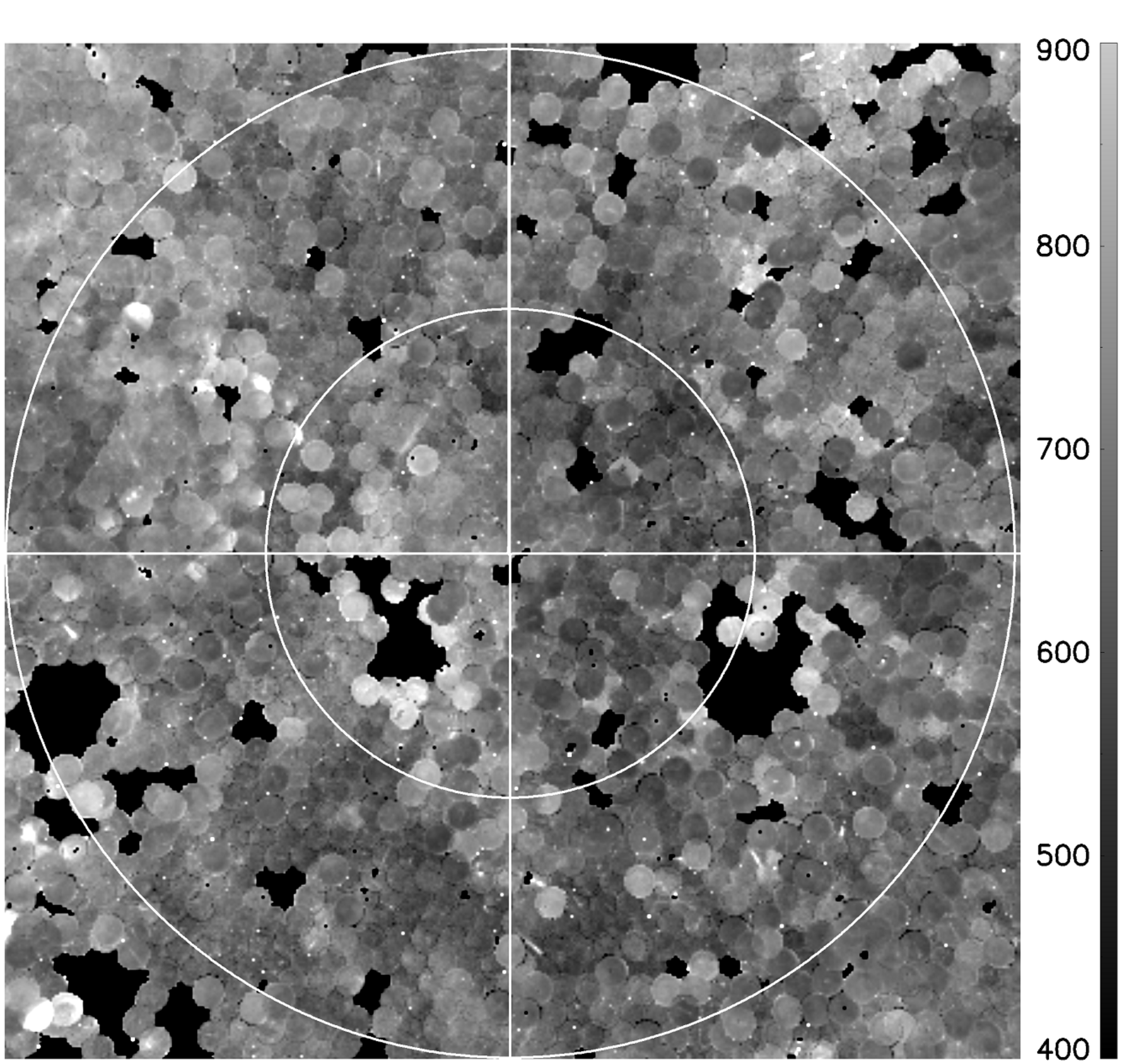}
    \includegraphics[scale=0.38]{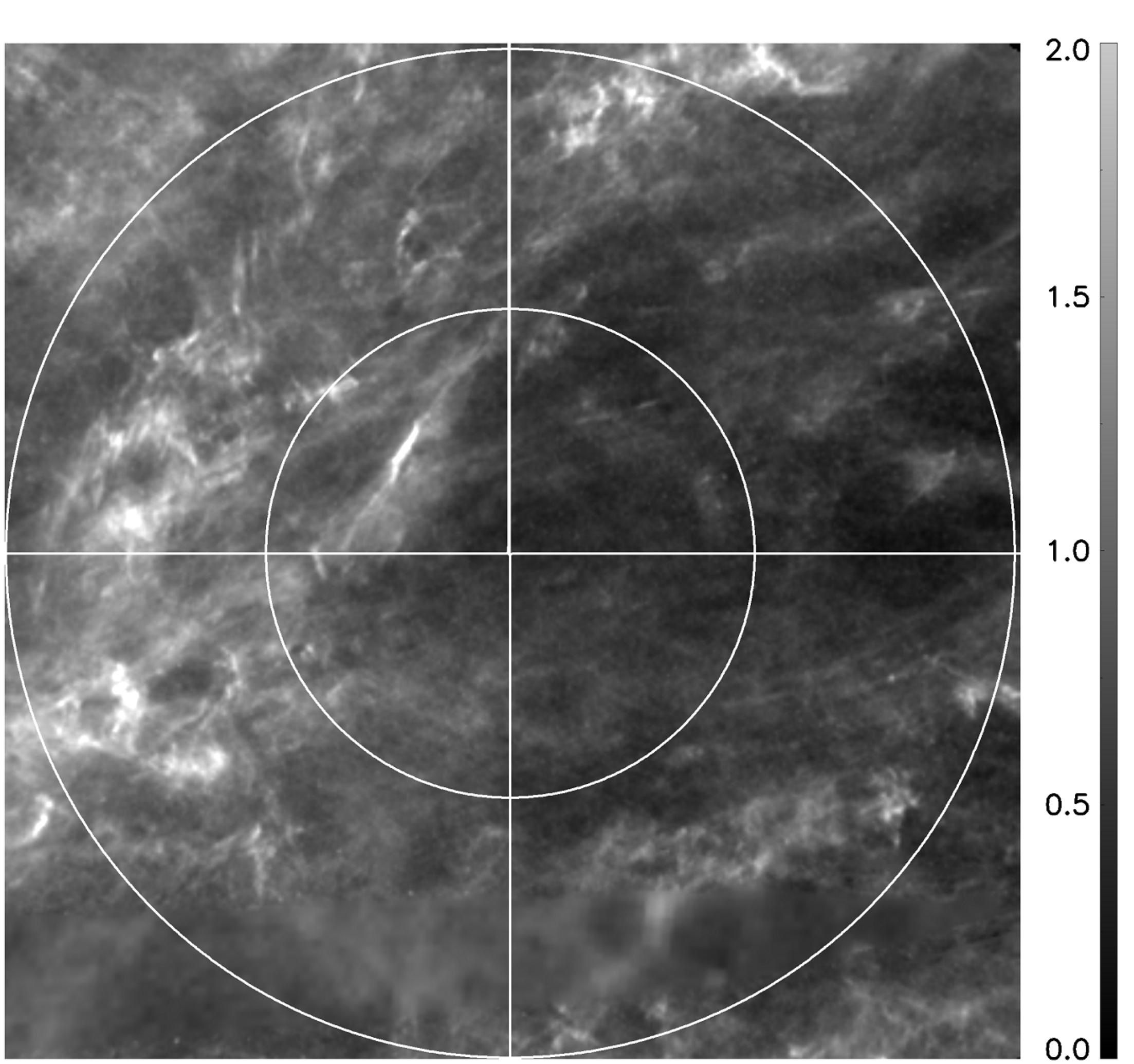}
    \includegraphics[scale=0.38]{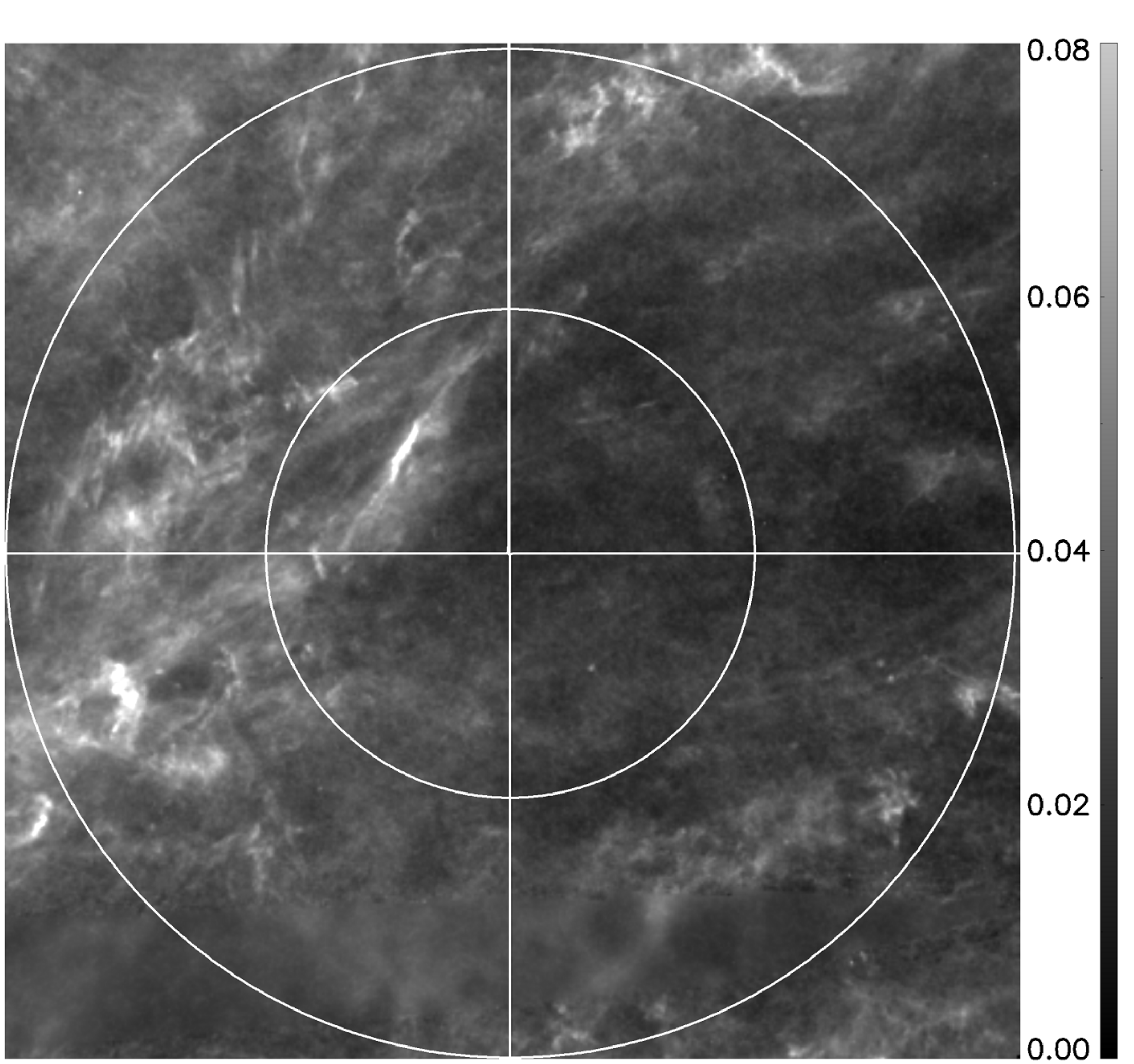}
   \caption{Diffuse FUV (top left), NUV (top right),  \citet{Schlegel1998} 100$\mu$m (bottom left) and \citet{Planck2014} extinction (bottom right) images are shown for the 70\degree$\leq$GLAT$\leq$90\degree\ region. The colour bars represent the surface brightness in \photu\ for the UV, in MJy sr$^{-1}$  for $100\mu$m, and magnitudes for the extinction map. The maps are at 6$'$ resolution. The Galactic longitudes are shown with 0\degree\ at the top and increasing 90\degree\ clockwise. The concentric circles indicate the Galactic latitudes 80\degree\ and 70\degree\ with 90\degree\ at the centre. The bright dots in the UV images represent artifacts around bright stars and have been rejected in our analysis.}
     \label{fig_maps}
 \end{figure*}

\begin{figure*}
	\includegraphics[scale=0.4]{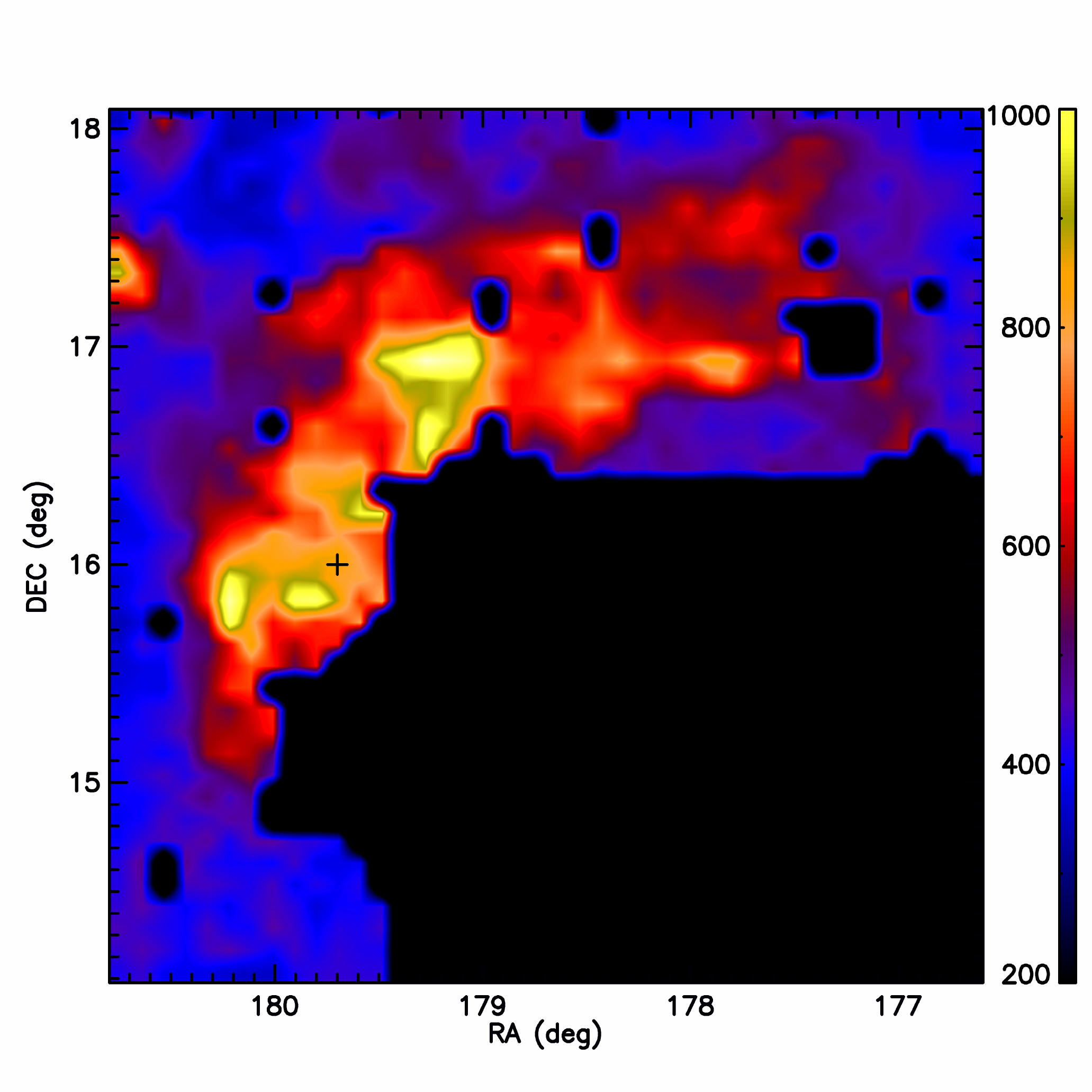}
    \includegraphics[scale=0.4]{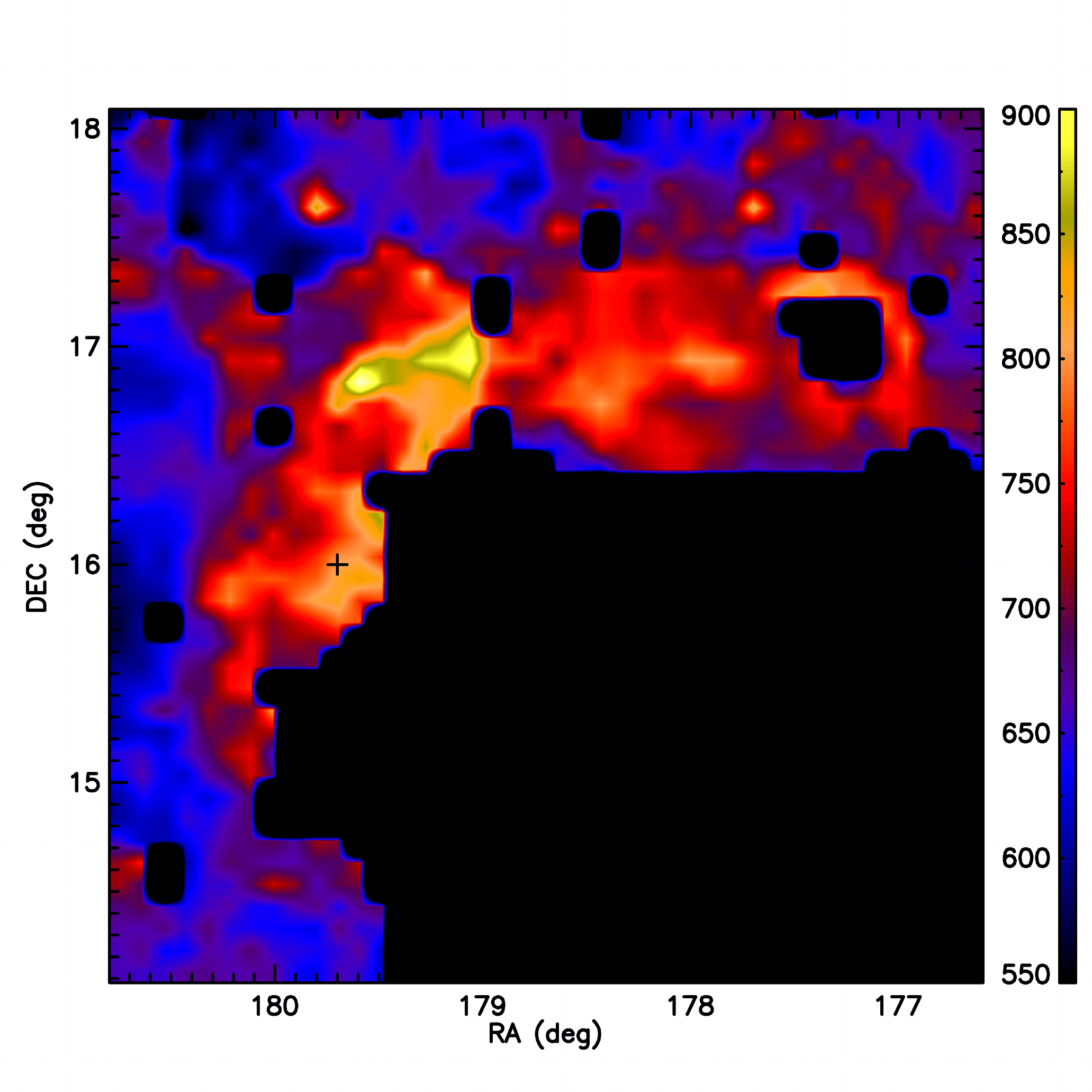}
    \caption{\galex\ diffuse FUV (left) and NUV (right) images of the Galactic cirrus cloud G251.2+73.3  (RA=179.7$\degree$, DEC=16$\degree$, marked with + symbol) studied by \citet{Haikala1995}. The colour bars represent the surface brightness in \photu. The field of view is 4$\degree\times\ $4$\degree$ with 6$'$ resolution.}
     \label{fig_cirrus_maps}
 \end{figure*}
 
We have chosen to study the region between 70\degree\ -- 80\degree\ in Galactic latitude (Fig. \ref{fig_maps}). The original motivation was to relook at a cirrus cloud discovered by \citet{Haikala1995} (Fig. \ref{fig_cirrus_maps}) using \textit{Far Ultraviolet Space Telescope} (FAUST: \citet{Bowyer1993}) data. These points occupy a limited range with 100\micron\ values between 1 -- 4 \mjy (0.2 < E(B - V) < 0.12 mag) and, based on a suggestion from an anonymous referee, we expanded our area of interest to the entire range between 70$\degree$ and 80$\degree$. The Haikala cloud may be seen to be part of much larger complex which includes Markkanen's Cloud \citep{Markkanen1979} near the North Galactic Pole (NGP). Cirrus clouds at high latitudes stand out against the general infrared background at 60 and 100\micron\ \citep{Low1984}. They scatter light from Galactic plane stars \citep{Jura1979} and block extragalactic radiation \citep{Mattila2017, Mattila2019}. Denser regions of the clouds may contain molecular hydrogen \citep{Weiland1986,Deul1990}, which emits in the FUV \citep{Martin1990,Hurwitz1998}. We have decomposed the different components of the diffuse UV radiation through a multi-wavelength analysis of the observed radiation at these latitudes. Although there is still not much interstellar dust at these latitudes, the peak column densities are higher than in the NGP and the effects of Galactic absorption and emission are more readily observed.

\section{Data}

\begin{figure*}
   \includegraphics[trim = 1.5cm 12.5cm 2cm 5cm, clip, scale=0.5]{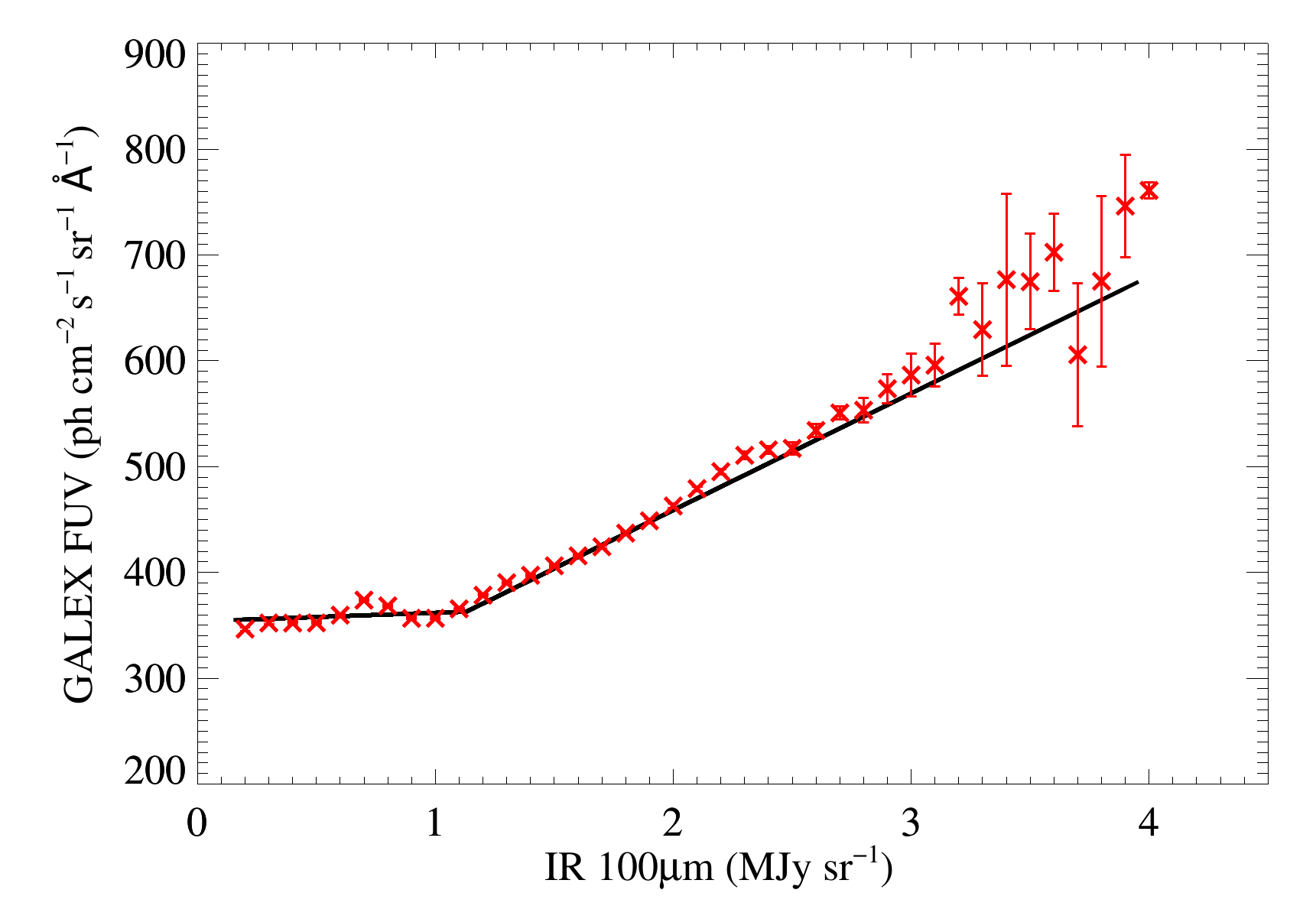}
   \includegraphics[trim = 1.5cm 12.5cm 2cm 5cm, clip, scale=0.5]{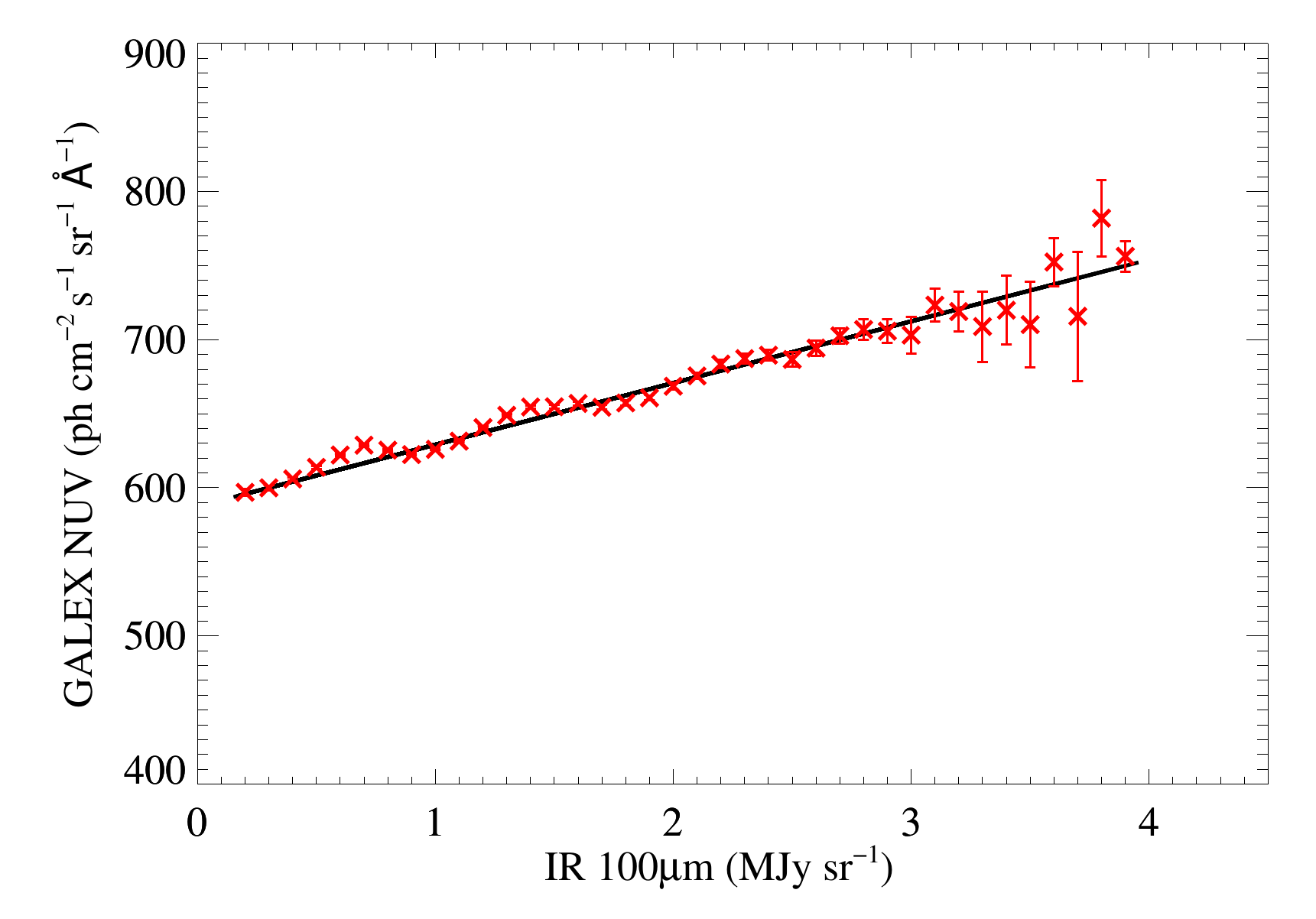}
   \includegraphics[trim = 1.5cm 12.5cm 2cm 5cm, clip, scale=0.5]{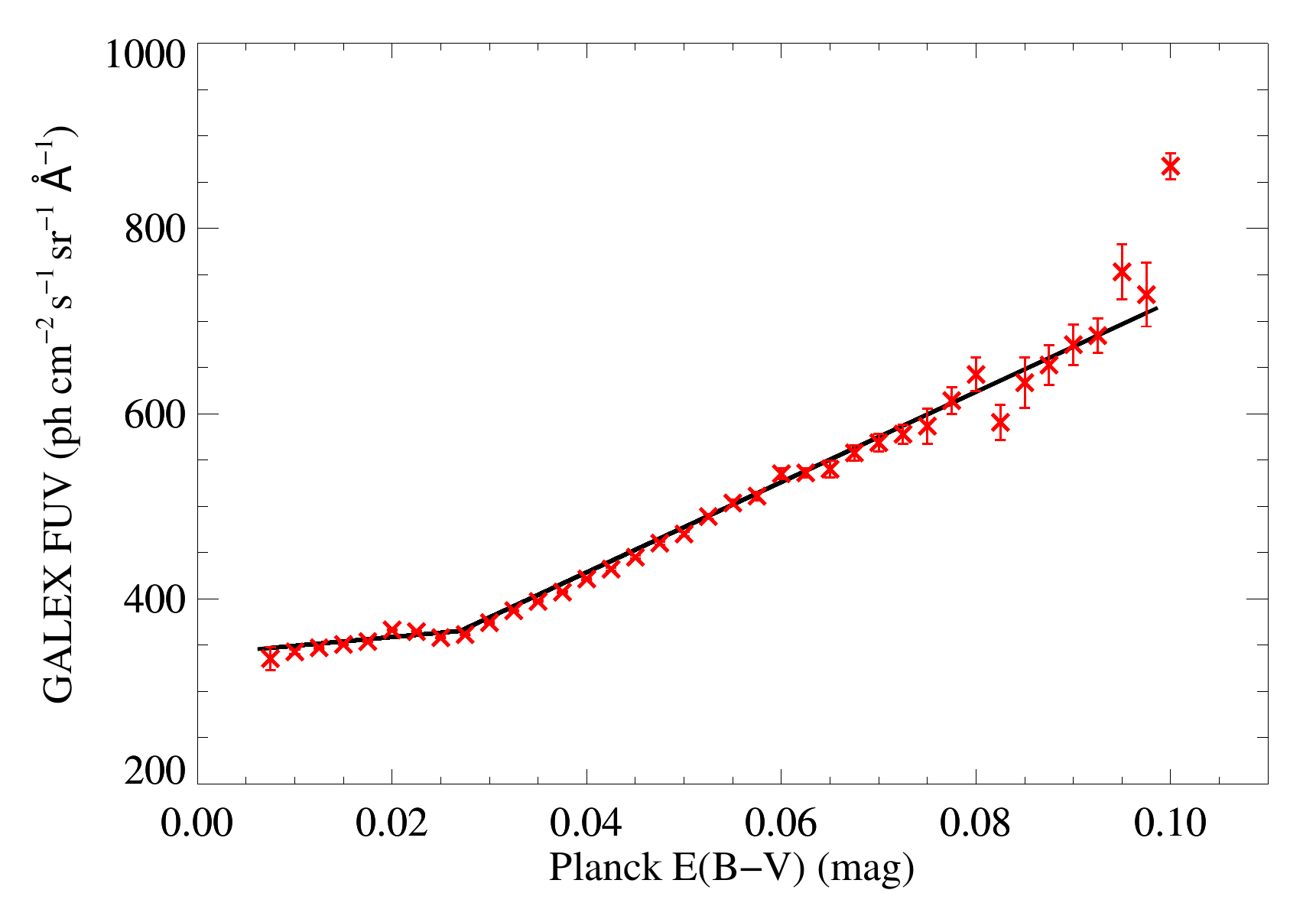}
   \includegraphics[trim = 1.5cm 12.5cm 2cm 5cm, clip, scale=0.5]{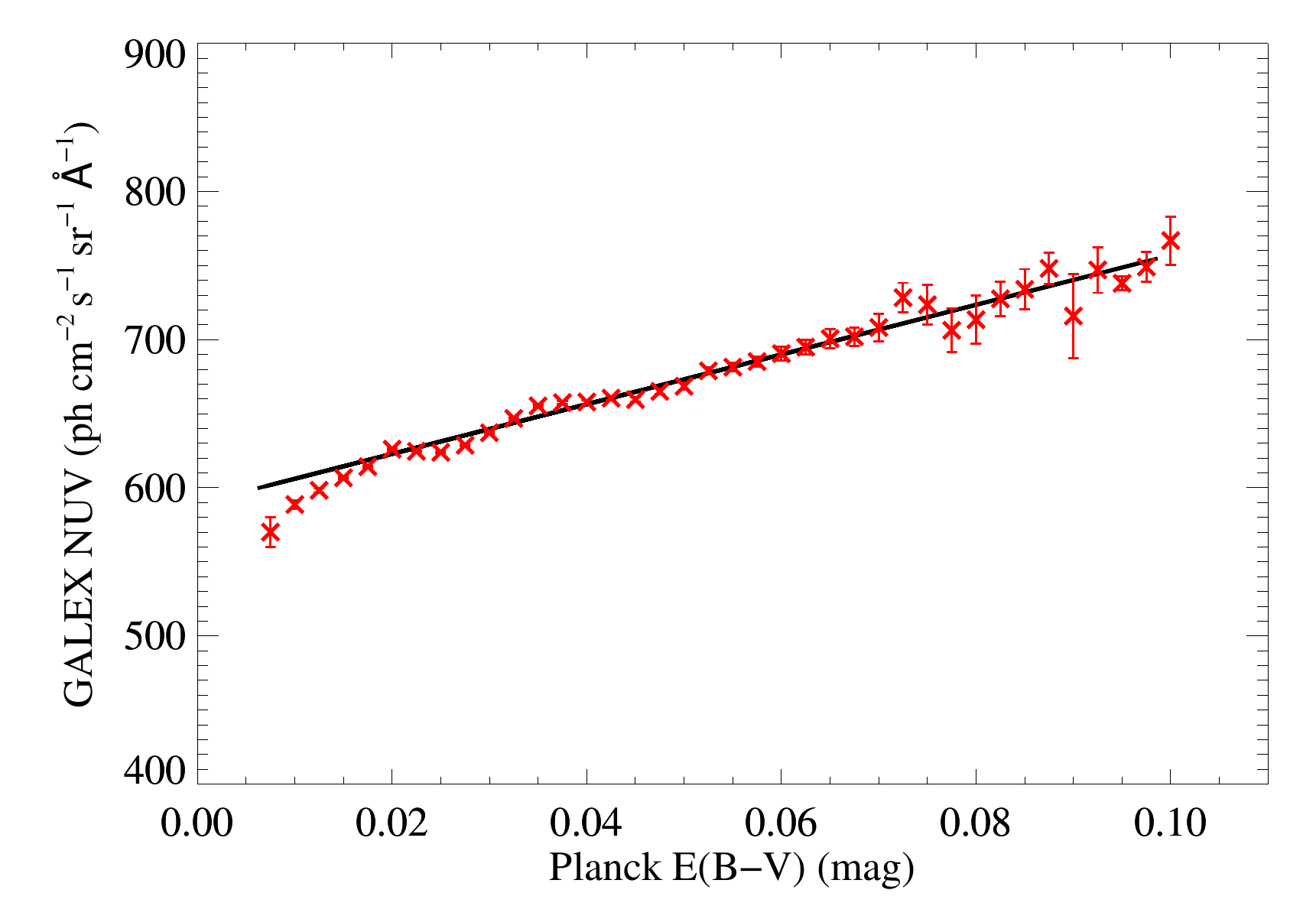}
   \caption{\emph{GALEX} FUV and NUV surface brightness in our selected region are plotted against 100$\mu$m surface brightness \citep{Schlegel1998} in the upper plots and against \citet{Planck2014} extinction values in the lower plots. Each point was calculated by averaging the UV surface brightness for a given IR or E(B-V) bin. The black lines indicate the best fits to the data. The standard error of the data points is overplotted in each plot.}
    \label{fig_uv_ir}
 \end{figure*}

We have collected archival data on the diffuse emission in this field from the UV to the IR. The UV data are from \citet{Murthy2014a} who extracted the diffuse emission in the NUV and FUV bands from \galex\ \citep{Martin2005}; the IR data at 100\micron\ from \citet{Schlegel1998} who used the data from the \emph{Infrared Astronomy Satellite} (\emph{IRAS}: \citet{Neugebauer1984}) and the \emph{Cosmic Background Explorer} (\emph{COBE}: \citet{Boggess1992}) to produce a reprocessed 100\micron\ all sky map; and the dust extinction from the $Planck$ mission \citep{Planck2014}. We have placed all the data on a common reference frame with 6$^{\prime}$ bins and displayed them in Fig. \ref{fig_maps}. The cirrus features in this field are distinctly seen in the \galex\ FUV, 100\micron\ and E(B-V) maps but not in the NUV, suggesting that dust scattering from the clouds is a small part of the total emission.

 \begin{figure}
   \includegraphics[trim = 1.5cm 12.5cm 2cm 5cm, clip, scale=0.5]{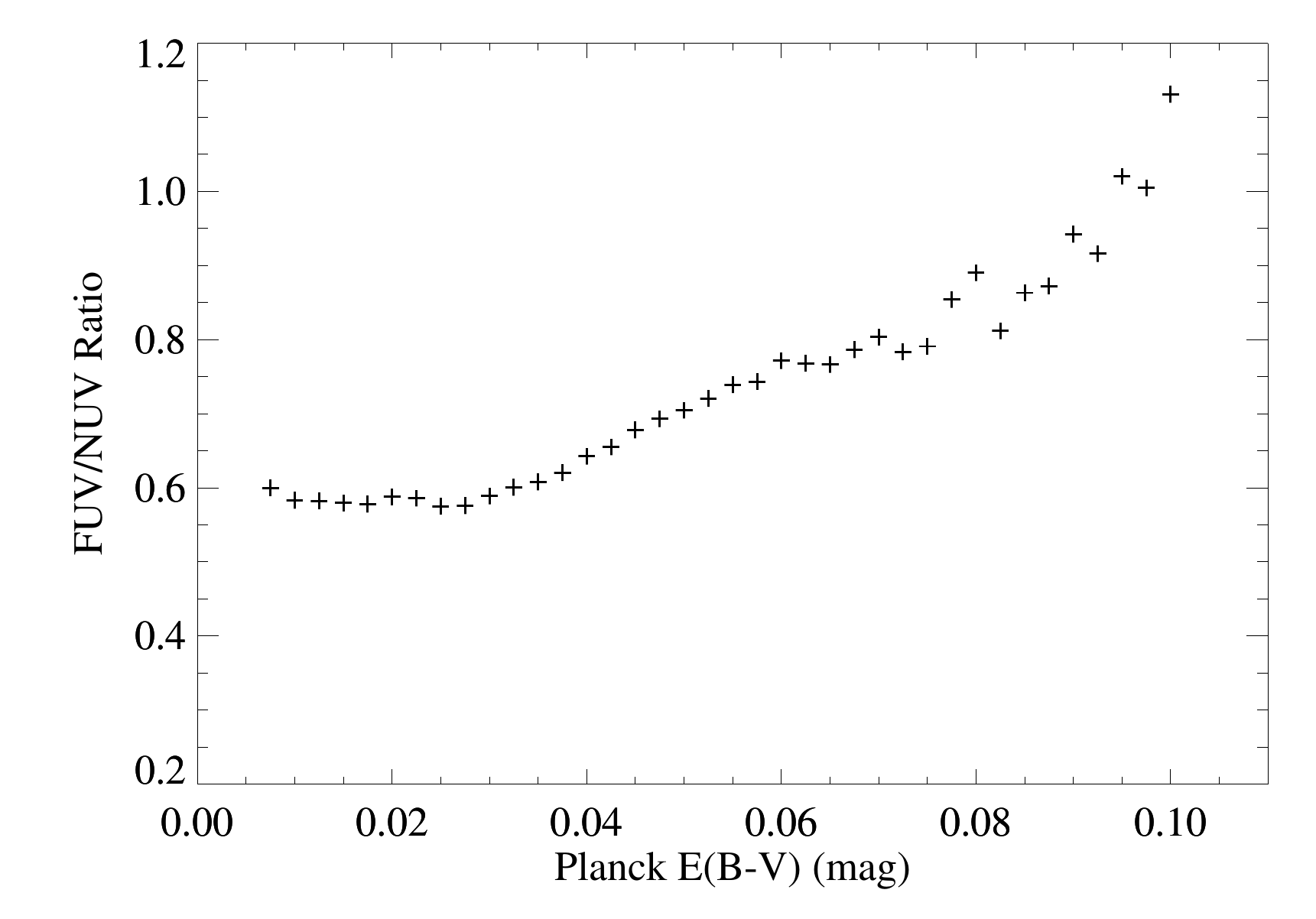}
   \caption{\galex\ FUV/NUV ratio plotted against E(B-V) by taking bins of 0.0025 mag along E(B-V). A raise in the ratio is observed close to the inflection point of 0.027 mag.}
    \label{fig_ratio_ebv}
 \end{figure}
 
 We have plotted the UV observations as a function of the IR and the E(B - V) in Fig. \ref{fig_uv_ir} and tabulated the correlations in Table \ref{tab1}. The NUV surface brightness is linearly correlated with both the 100\micron\ and the E(B -V). The FUV surface brightness is also correlated with the IR and the E(B - V) but with an inflection point at $1.13 \pm 0.13$ \mjy\ (0.027 mag in E(B - V)) where the slope increases. This inflection point is also seen in the FUV/NUV ratio (Fig. \ref{fig_ratio_ebv}). The UV radiation and the IR are both due to the amount of dust along the line of sight and one might expect them all to be tightly correlated, especially at low column densities where the optical depth is low both in the UV and the IR. The excess radiation in the FUV is most likely due to Lyman band fluorescent emission from molecular hydrogen \citep{Martin1990,Hurwitz1998}. The inflection point represents the column density above which self-shielding protects the hydrogen molecules from photodissociation by interstellar UV photons and corresponds to a column density of log N(H) = 20.2.  Canonically, this critical column density has been taken as log N(H)$>$20.6 \citep{Savage1977,Franco1986,Reach1994} but has been found to be lower in high latitudes cirrus clouds \citep{Gillmon2006,Planck2011}, where \citet{Gillmon2006b} suggests that H$_2$ formation is more efficient due to compression from dynamical processes at the disk-halo interface.
 
 The intercepts observed represents the surface brightness at zero column density where we would expect no contribution from dust scattering. We will discuss its possible components below but note that they are in agreement with the spectroscopic data of \citet{Anderson1979, Tennyson1988, Seon2011} who found similar offsets after correcting for airglow lines in the spectrum.

\begin{table}
\centering
\caption{Correlation coefficients}
\label{tab1}
\begin{tabular}{lllll}
\hline \hline
Wavelength & r & a & b & $\chi^2$ \\ \hline
FUV--100$\mu$m ($<$1.13 \mjy) &  0.03 &   7.97 &  353.89 &    0.84 \\
FUV--100$\mu$m ($\geq$1.13 \mjy) &  0.46 &  110.35 &   238.20 &    0.70 \\
FUV--E(B-V) ($<$ 0.027 mag) &  0.06 & 942.14 &  339.97 &    0.82 \\
FUV--E(B-V) ($\geq$ 0.027 mag) &  0.47 & 4871.51 &  233.88 &    0.69 \\
NUV--100$\mu$m & 0.25 &   41.63 &  587.47 &    0.94 \\
NUV--E(B-V) &  0.27 & 1677.25 &  589.33 &    0.52 \\ \hline
\hline
\multicolumn{5}{c}{NGP$^{*}$} \\ \hline
FUV--100$\mu$m ($<$1.08 \mjy) & 0.27 & 57.43 & 288.27 & 1.15 \\
FUV--100$\mu$m ($>$1.08 \mjy) & 0.57 & 156.33 & 182.10 & 1.37 \\
FUV--E(B-V) & 0.52 & 4245.40 & 250.11 & 1.24 \\
NUV--100$\mu$m & 0.42 & 68 & 530.89 & 1.18 \\
NUV--E(B-V) & 0.40 & 2655.67 & 531.06 & 1.21 \\ \hline 
\multicolumn{5}{l}{r - Spearman's correlation coefficient (P $<<<$ 0.05 for all cases)} \\
\multicolumn{5}{l}{a - Scale factor} \\
\multicolumn{5}{l}{b - Offset (\photu)} \\
\multicolumn{5}{l}{$^{*}$ From \citet{Akshaya2018}}
\end{tabular}
\end{table}

\section{Modelling and Results}

 We have assumed four independent sources for the UV radiation in this field:
\begin{enumerate}
\item Extragalactic background light (EBL).
\item Starlight scattered from interstellar dust.
\item Molecular hydrogen fluorescence.
\item Unexplained offset.
\end{enumerate}
We will discuss each of these in the following sections.

\subsection{EBL}

\begin{figure}
   \includegraphics[trim = 1.5cm 12.5cm 2cm 4cm, clip, scale=0.5]{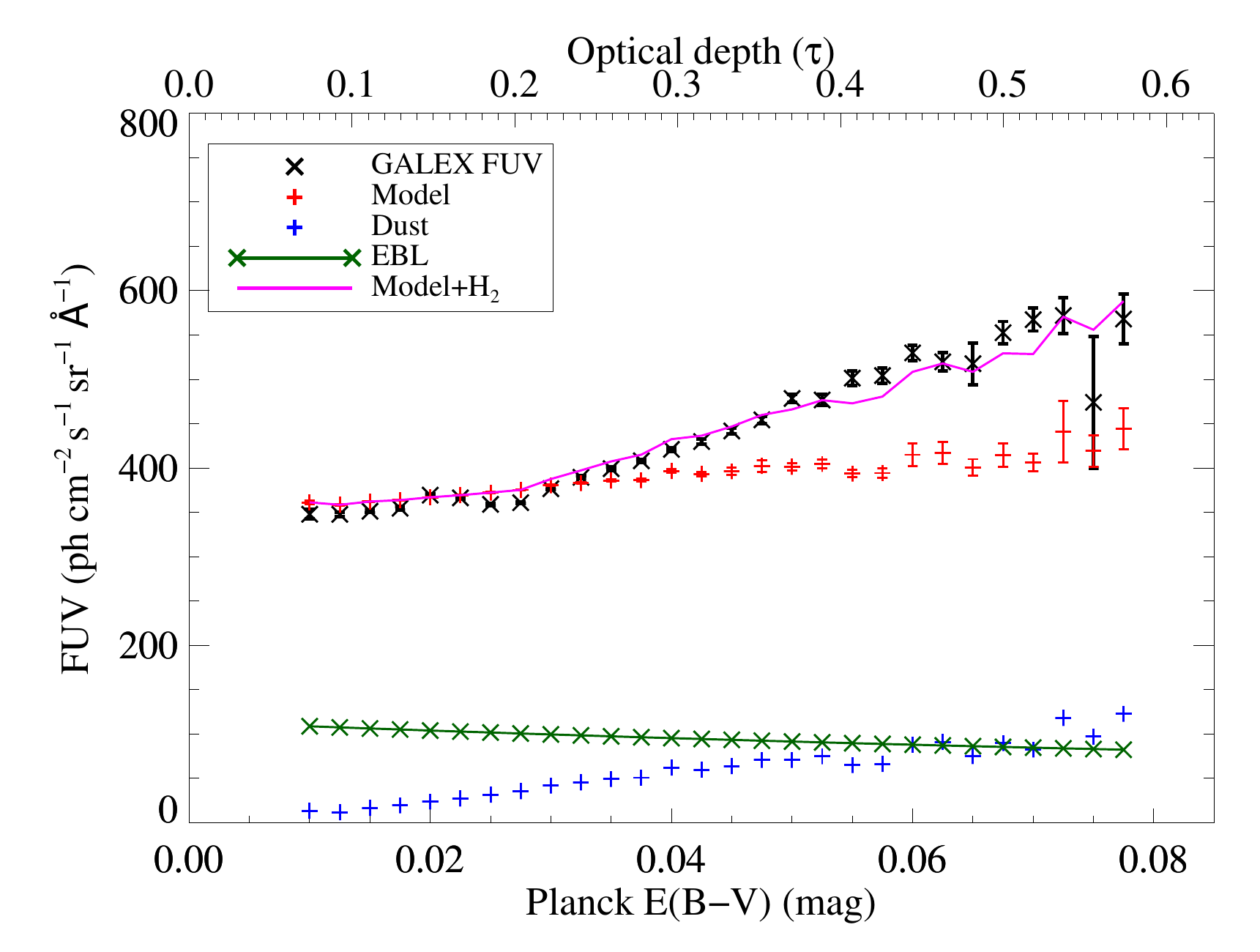}
   \includegraphics[trim = 1.5cm 12.5cm 2cm 4cm, clip, scale=0.5]{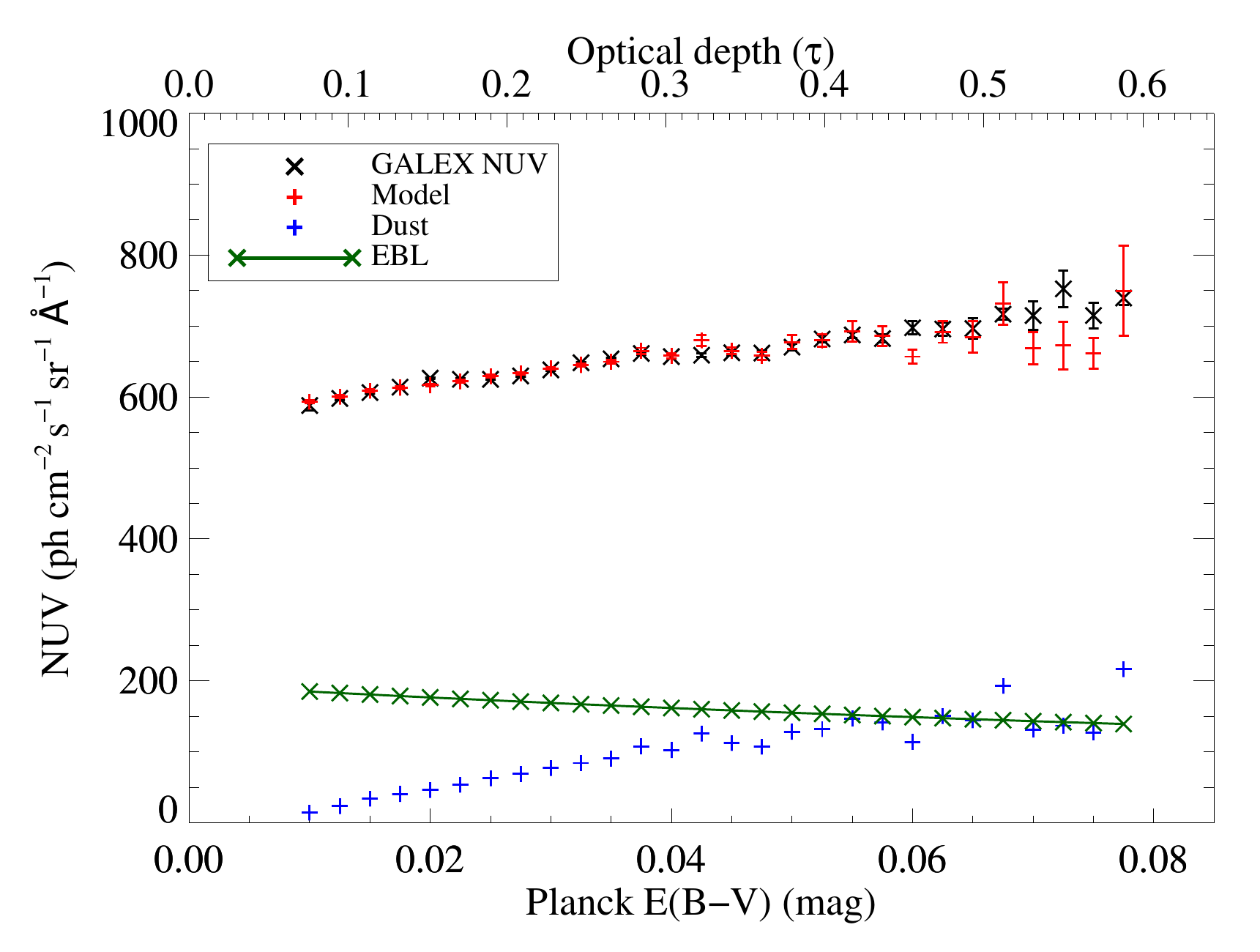}
   \caption{Modelled and observed UV surface brightness plotted against the \citet{Planck2014} E(B-V). The modelled surface brightness includes contributions from the EBL (green X), the dust scattered light (blue +) and the zero-offset. Fluorescent emission from H$_{2}$ is added to the other components for the FUV. Standard errors are overplotted for both the observed data and the model. We have only considered those bins with more than five data points to compare with the model.}
    \label{fig_model_fits}
 \end{figure}

\citet{Akshaya2018} tabulated the different components of the known EBL:
\begin{itemize}
    \item 60 -- 81 \photu\ in FUV and 121 -- 181 \photu\ in NUV from the integrated light from galaxies \citep{Gardner2000,Xu2005,Voyer2011,Driver2016};
    \item 16 -- 30 \photu\ from QSOs \citep{Madau1992};
    \item $<$20 \photu\ from the intergalactic medium (IGM) \citep{Martin1991};
\end{itemize}
for a total of $114 \pm 18$ \photu\ in the FUV and $194 \pm 37$ \photu\ in the NUV. We have extincted the EBL using the known E(B - V) and accounted for the scattered EBL photons for different values of albedo and phase function asymmetry factor using \citet{Mattila1976}. This is shown in Fig. \ref{fig_model_fits} (green line).

\begin{table}
\centering
\caption{Albedo and phase function asymmetry factor from literature}
\label{tab_ag}
\begin{tabular}{p{3.64cm}p{1.4cm}p{1.22cm}l}
\hline \hline
References & Wavelength (\AA) & $a$ & $g$ \\ \hline
\citet{Onaka1991} & 1450--1800 & $>$0.32 & $>$0.5 \\
\citet{Henry1993} & 1500 & $>0.5$ & $>0.7$ \\
\citet{Witt1993} & 1000--1600 & 0.42$\pm$0.04 & 0.75 \\
\citet{Murthy1993b} & 912--1150 & $>0.3$ & $<$0.8 \\
\citet{Gordon1994} & 1362 & 0.47--0.7 & $<$0.8 \\
 & 1769 & 0.55--0.73 & $<$0.8 \\
\citet{Witt1994} & 1500 & 0.5 & 0.9 \\
\citet{Sasseen1996} & 1400--1800 & 0.3 & 0.8 \\
\citet{Witt1997} & 1400--1800 & 0.45$\pm$0.05 & 0.68$\pm$0.1 \\
\citet{Schiminovich2001} & 1740 & 0.45$\pm$0.05 & 0.77$\pm$0.1 \\
\citet{Burg2002} & 900--1400 & 0.2--0.4 & 0.85 \\
\citet{Mathis2002} & 1300 & $\ge$0.5 & 0.6--0.85 \\
\citet{Shalima2004} & 1100 & 0.4$\pm$0.2 & - \\
\citet{Sujatha2005} & 1100 & 0.4$\pm$0.1 & 0.55$\pm$0.25 \\
\citet{Shalima2006} & 900--1200 & 0.3--0.7 & 0.55--0.85 \\
\citet{Lee2008} & 1370--1670 & 0.36$\pm$0.2 & 0.52$\pm$0.22 \\
\citet{Sujatha2009} & 1350--1750 & 0.4 & 0.7 \\
\citet{Puthiyaveettil2010} & 1400--1900 & 0.6 & 0.8 \\
\citet{Sujatha2010} & 1350--1750 & 0.32$\pm$0.09 & 0.51$\pm$0.19 \\
 & 1750--2850 & 0.45$\pm$0.08 & 0.56$\pm$0.1 \\
\citet{Jo2012} & 1350--1750 & 0.42$\pm$0.03 & 0.45$\pm$0.02 \\
\citet{Choi2013} & 1330--1780 & 0.38$\pm$0.06 & 0.46$\pm$0.06 \\
\citet{Hamden2013} & 1344--1786 & 0.62$\pm$0.04 & 0.78$\pm$0.05 \\
\citet{Lim2013} & 1360--1680 & 0.42$\pm$0.05 & 0.2--0.58 \\
\citet{Murthy2016} & 1500 & 0.4 & 0.6 \\
 & 2300 & 0.4 & 0.6 \\
\citet{Mattila2018} & 3500 & 0.58$\pm$0.05 & 0.6$\pm$0.1 \\
 \hline
\end{tabular}
\end{table}

\subsection{Dust-scattered Starlight}
We have used the same Monte Carlo model for the dust-scattered starlight as did \citet{Akshaya2018}. This model is fully described in \citet{Murthy2016} and may be downloaded from \citet{MurthyANSIC}. The model uses the known positions of hot, UV stars from the Hipparcos star catalog \citep{Perryman1997} with model spectra from \citet{Castelli2003} to predict the number of photons at any position in the Galaxy. The amount and location of the interstellar dust was derived from \citet{Schlegel1998} reddening maps using a scale height of 125pc \citep{Marshall2006}. The stellar photons were scattered from the dust using the Henyey-Greenstein scattering function \citep{Henyey1941} and assuming a range of values for albedo ($a$) and phase function asymmetry factor ($g$). 

We derived the dust scattered light by subtracting the zero-offset (Table \ref{tab1}) and EBL with extinction from the total observed radiation. We then fit this remainder with the model predictions for different values of $a$ and $g$ for the entire data in NUV and for E(B-V)$<$0.027 in case of FUV. As there is lot of scatter in the model they were compared with the data at a resolution of 12$'$. We have only considered those bins with more than five data points to compare with the model. We find 1$\sigma$ confidence interval of $a=0.4\pm0.1$ and $g=0.8\pm0.1$ in the FUV and $a=0.4\pm0.1$ and $g=0.5\pm0.1$ in the NUV using the formulation given by \citet{Lampton1976}. These are in agreement with the theoretical predictions of \citet{Draine2003} and various observational estimates (Table \ref{tab_ag}). The models fit the data well in the NUV but fall short in the FUV for E(B - V) $>$ 0.027, where we believe that the fluorescent emission from molecular hydrogen begins to be important.

\subsection{Molecular Hydrogen Fluorescence}
\begin{figure}
   \includegraphics[trim = 1.5cm 12.5cm 2cm 5cm, clip, scale=0.5]{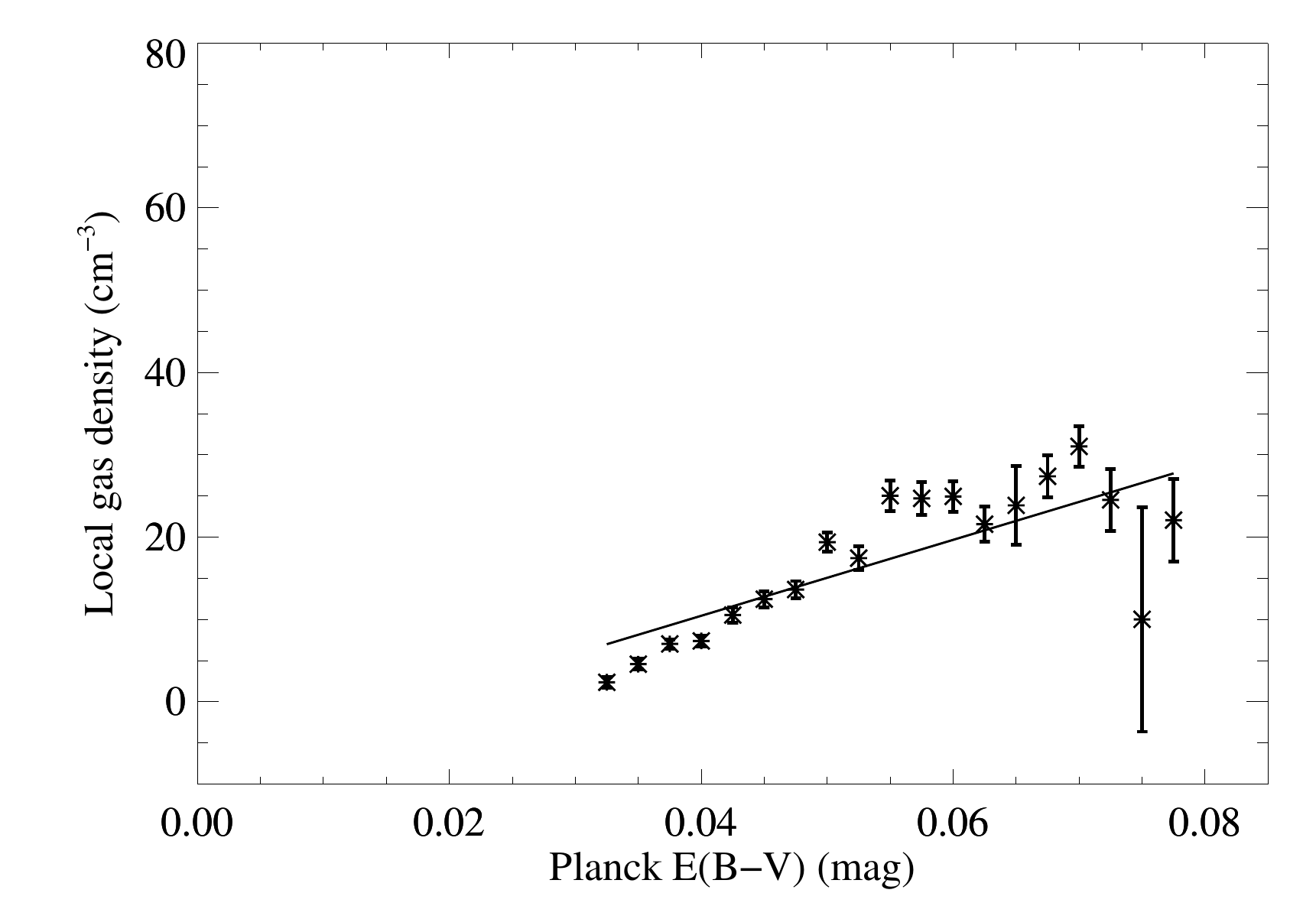}
   \caption{The gas density in the cloud (assuming $\overline{n\textrm{(H{\scriptsize I})}} \approx n_{\textrm{H}}$) at different values of E(B-V) at the regions where fluorescence is observed, obtained from the modelling presented by \citet{Martin1990}. The black line shows the best fit to the data.}
    \label{fig_nh_vol}
 \end{figure}

We have plotted the contributions to the observed data from the EBL, the dust-scattered starlight and the offset in Fig. \ref{fig_model_fits}. The NUV data are fit well by the model as are the FUV data for E(B - V) $<$ 0.3, corresponding to log N(H) = 20.2, assuming N(H)/E(B-V) = 5.8$\times$10$^{21}$ atoms cm$^{-2}$ mag$^{-1}$  \citep{Bohlin1978}. We have attributed this excess emission to fluorescence in the Lyman band (1400 -- 1700 \AA) of molecular hydrogen \citep{Duley1980,Martin1990,Hurwitz1998}. The level of this emission averages out to about 100 \photu, consistent with the surface brightness found by \citet{Jo2017}. Using the formulation of \citet{Martin1990}, we can derive the local gas density in the cloud from our observations assuming a constant value of R = $1\times10^{-17}$ cm$^{3}$ s$^{-1}$ for the formation rate of H$_{2}$ (Fig. \ref{fig_nh_vol}).  Adding this emission to the other components gives the solid line in Fig. \ref{fig_model_fits}.

 \subsection{Offsets}

It has long been noted \citep{Henry1991,Hamden2013, Henry2015} that the diffuse background, particularly at high latitudes, cannot be fully explained by known sources (dust scattered radiation and EBL) without an additional offset at zero column density. \citet{Akshaya2018} found this offset to be $174 \pm 18$ (FUV) and $337 \pm 37$ (NUV) \photu\ in the North Galactic Pole and $127 \pm 18$ (FUV) and $385 \pm 37$ (NUV) \photu\ in the South Galactic Pole. In this work, we find that the offsets are $240 \pm 18$ and $394 \pm 37$ \photu\ in the FUV and NUV, respectively, after subtracting the known EBL. These offsets are slightly higher than those in the polar regions which may be due to uncertainties in the data. 

Unfortunately, we cannot distinguish residual airglow from other diffuse sources in our study but other, spectroscopic, studies have found similar offsets at the NGP \citep{Anderson1979, Tennyson1988, Seon2011}. \citet{Brune1978} has measured the contribution from the 1356 \AA\ OI line, the only significant contributor in the \textit{GALEX} FUV band, to be less than 50 photon units. We are therefore confident that the offset is predominantly not due to airglow.

 \section{Conclusions}

We have studied the latitude range between 70\degree\ and 80\degree\ as a continuation of our characterisation of the diffuse UV sky. This region has a greater amount of nebulosity and hence higher column densities than our earlier study at the Galactic Poles \citep{Akshaya2018}. We confirm the presence of an offset at zero column density (E(B -V) = 0) with a level of $240 \pm 18$ \photu\ in the FUV and $394 \pm 37$ \photu\ in the NUV, slightly higher than in the polar regions \citep{Akshaya2018}. \citet{Henry2015} has suggested that this is due to an unknown Galactic component of the DGL but it may be due to an unidentified EBL component. This may be resolved as we extend our study to lower Galactic latitudes where the dust component is enhanced and the extragalactic component is further extincted.

We find a component due to scattering from interstellar dust rising to a level of about 100 \photu\ in the FUV and 200 \photu\ in the NUV, about the same level as the EBL. We were able to set 1$\sigma$ limits on $a$ to be 0.4$\pm$0.1 in both FUV and NUV and the limits on $g$ were found to be $g=0.8\pm0.1$ in FUV and $g=0.5\pm0.1$ in NUV. We find an increase over the dust scattered emission in the FUV for column densities greater than E(B - V) = 0.027 mag (log N(H) = 20.2) which we interpret as molecular hydrogen fluorescent emission in the Lyman band.

Molecular hydrogen fluorescence emission was observed in the NGP by \citet{Akshaya2018} for column densities greater than log N(H)=20.2. We also observe this emission at the same column density for the region 70\degree$\leq$GLAT$\leq80$\degree. Other studies have found a higher value (log N(H) $>$ 20.6) for the critical density at which molecular hydrogen can form through self-shielding from the interstellar UV photons \citep{Savage1977,Franco1986,Reach1994} but \citet{Gillmon2006b} suggest that the critical density is less at high Galactic latitudes.

\section*{Acknowledgements}

Support for JM has come from proposal EMR/2016/001450 from DST/SERB.
We have used the Gnu Data Language (http://gnudatalanguage.sourceforge.net/index.php) for the analysis of these data. The data presented in this paper were obtained from the Mikulski Archive for Space Telescopes (MAST). STScI is operated by the Association of Universities for research in Astronomy, Inc., under the NASA contract NAS5-26555.  Support for MAST for non-HST data is provided by the NASA Office of Space Science via grant NNX09AF08G and by other grants and contracts.




\bibliographystyle{mnras}
\bibliography{ref} 

\begin{thebibliography}{}
\makeatletter
\relax
\def\mn@urlcharsother{\let\do\@makeother \do\$\do\&\do\#\do\^\do\_\do\%\do\~}
\def\mn@doi{\begingroup\mn@urlcharsother \@ifnextchar [ {\mn@doi@}
  {\mn@doi@[]}}
\def\mn@doi@[#1]#2{\def\@tempa{#1}\ifx\@tempa\@empty \href
  {http://dx.doi.org/#2} {doi:#2}\else \href {http://dx.doi.org/#2} {#1}\fi
  \endgroup}
\def\mn@eprint#1#2{\mn@eprint@#1:#2::\@nil}
\def\mn@eprint@arXiv#1{\href {http://arxiv.org/abs/#1} {{\tt arXiv:#1}}}
\def\mn@eprint@dblp#1{\href {http://dblp.uni-trier.de/rec/bibtex/#1.xml}
  {dblp:#1}}
\def\mn@eprint@#1:#2:#3:#4\@nil{\def\@tempa {#1}\def\@tempb {#2}\def\@tempc
  {#3}\ifx \@tempc \@empty \let \@tempc \@tempb \let \@tempb \@tempa \fi \ifx
  \@tempb \@empty \def\@tempb {arXiv}\fi \@ifundefined
  {mn@eprint@\@tempb}{\@tempb:\@tempc}{\expandafter \expandafter \csname
  mn@eprint@\@tempb\endcsname \expandafter{\@tempc}}}

\bibitem[\protect\citeauthoryear{{Akshaya}, {Murthy}, {Ravichandran}, {Henry}
  \& {Overduin}}{{Akshaya} et~al.}{2018}]{Akshaya2018}
{Akshaya} M.~S.,  {Murthy} J.,  {Ravichandran} S.,  {Henry} R.~C.,   {Overduin}
  J.,  2018, \mn@doi [\apj] {10.3847/1538-4357/aabcb9}, \href
  {http://adsabs.harvard.edu/abs/2018ApJ...858..101A} {858, 101}

\bibitem[\protect\citeauthoryear{{Anderson}, {Henry}, {Brune}, {Feldman}  \&
  {Fastie}}{{Anderson} et~al.}{1979}]{Anderson1979}
{Anderson} R.~C.,  {Henry} R.~C.,  {Brune} W.~H.,  {Feldman} P.~D.,   {Fastie}
  W.~G.,  1979, \mn@doi [\apj] {10.1086/157510}, \href
  {http://adsabs.harvard.edu/abs/1979ApJ...234..415A} {234, 415}

\bibitem[\protect\citeauthoryear{{Boggess} et~al.,}{{Boggess}
  et~al.}{1992}]{Boggess1992}
{Boggess} N.~W.,  et~al., 1992, \mn@doi [\apj] {10.1086/171797}, \href
  {https://ui.adsabs.harvard.edu/abs/1992ApJ...397..420B} {397, 420}

\bibitem[\protect\citeauthoryear{{Bohlin}, {Savage}  \& {Drake}}{{Bohlin}
  et~al.}{1978}]{Bohlin1978}
{Bohlin} R.~C.,  {Savage} B.~D.,   {Drake} J.~F.,  1978, \mn@doi [\apj]
  {10.1086/156357}, \href {http://cdsads.u-strasbg.fr/abs/1978ApJ...224..132B}
  {224, 132}

\bibitem[\protect\citeauthoryear{{Boissier} et~al.,}{{Boissier}
  et~al.}{2015}]{Boissier2015}
{Boissier} S.,  et~al., 2015, \mn@doi [\aap] {10.1051/0004-6361/201526089},
  \href {http://adsabs.harvard.edu/abs/2015A%26A...579A..29B} {579, A29}

\bibitem[\protect\citeauthoryear{{Bowyer}}{{Bowyer}}{1991}]{Bowyer1991}
{Bowyer} S.,  1991, \mn@doi [\araa] {10.1146/annurev.aa.29.090191.000423},
  \href {http://cdsads.u-strasbg.fr/abs/1991ARA%26A..29...59B} {29, 59}

\bibitem[\protect\citeauthoryear{{Bowyer}, {Sasseen}, {Lampton}  \&
  {Wu}}{{Bowyer} et~al.}{1993}]{Bowyer1993}
{Bowyer} S.,  {Sasseen} T.~P.,  {Lampton} M.,   {Wu} X.,  1993, \mn@doi [\apj]
  {10.1086/173209}, \href {http://adsabs.harvard.edu/abs/1993ApJ...415..875B}
  {415, 875}

\bibitem[\protect\citeauthoryear{{Brune}, {Feldman}, {Anderson}, {Fastie}  \&
  {Henry}}{{Brune} et~al.}{1978}]{Brune1978}
{Brune} W.~H.,  {Feldman} P.~D.,  {Anderson} R.~C.,  {Fastie} W.~G.,   {Henry}
  R.~C.,  1978, \mn@doi [\grl] {10.1029/GL005i005p00383}, \href
  {https://ui.adsabs.harvard.edu/abs/1978GeoRL...5..383B} {5, 383}

\bibitem[\protect\citeauthoryear{{Burgh}, {McCandliss}  \& {Feldman}}{{Burgh}
  et~al.}{2002}]{Burg2002}
{Burgh} E.~B.,  {McCandliss} S.~R.,   {Feldman} P.~D.,  2002, \mn@doi [\apj]
  {10.1086/341194}, \href
  {https://ui.adsabs.harvard.edu/abs/2002ApJ...575..240B} {575, 240}

\bibitem[\protect\citeauthoryear{{Castelli} \& {Kurucz}}{{Castelli} \&
  {Kurucz}}{2003}]{Castelli2003}
{Castelli} F.,  {Kurucz} R.~L.,  2003, in {Piskunov} N.,  {Weiss} W.~W.,
  {Gray} D.~F.,  eds,  IAU Symposium Vol. 210, Modelling of Stellar
  Atmospheres. p.~A20 (\mn@eprint {} {astro-ph/0405087})

\bibitem[\protect\citeauthoryear{{Choi}, {Min}, {Seon}, {Lim}, {Jo}  \&
  {Park}}{{Choi} et~al.}{2013}]{Choi2013}
{Choi} Y.-J.,  {Min} K.-W.,  {Seon} K.-I.,  {Lim} T.-H.,  {Jo} Y.-S.,   {Park}
  J.-W.,  2013, \mn@doi [\apj] {10.1088/0004-637X/774/1/34}, \href
  {https://ui.adsabs.harvard.edu/abs/2013ApJ...774...34C} {774, 34}

\bibitem[\protect\citeauthoryear{{Deul} \& {Burton}}{{Deul} \&
  {Burton}}{1990}]{Deul1990}
{Deul} E.~R.,  {Burton} W.~B.,  1990, \aap, \href
  {https://ui.adsabs.harvard.edu/\#abs/1990A&A...230..153D} {230, 153}

\bibitem[\protect\citeauthoryear{{Draine}}{{Draine}}{2003}]{Draine2003}
{Draine} B.~T.,  2003, \mn@doi [\araa]
  {10.1146/annurev.astro.41.011802.094840}, \href
  {http://adsabs.harvard.edu/abs/2003ARA%26A..41..241D} {41, 241}

\bibitem[\protect\citeauthoryear{{Driver} et~al.,}{{Driver}
  et~al.}{2016}]{Driver2016}
{Driver} S.~P.,  et~al., 2016, \mn@doi [\apj] {10.3847/0004-637X/827/2/108},
  \href {http://adsabs.harvard.edu/abs/2016ApJ...827..108D} {827, 108}

\bibitem[\protect\citeauthoryear{{Duley} \& {Williams}}{{Duley} \&
  {Williams}}{1980}]{Duley1980}
{Duley} W.~W.,  {Williams} D.~A.,  1980, \mn@doi [\apjl] {10.1086/183427},
  \href {http://adsabs.harvard.edu/abs/1980ApJ...242L.179D} {242, L179}

\bibitem[\protect\citeauthoryear{{Franco} \& {Cox}}{{Franco} \&
  {Cox}}{1986}]{Franco1986}
{Franco} J.,  {Cox} D.~P.,  1986, \mn@doi [\pasp] {10.1086/131876}, \href
  {http://adsabs.harvard.edu/abs/1986PASP...98.1076F} {98, 1076}

\bibitem[\protect\citeauthoryear{{Gardner}, {Brown}  \& {Ferguson}}{{Gardner}
  et~al.}{2000}]{Gardner2000}
{Gardner} J.~P.,  {Brown} T.~M.,   {Ferguson} H.~C.,  2000, \mn@doi [\apjl]
  {10.1086/312930}, \href {http://adsabs.harvard.edu/abs/2000ApJ...542L..79G}
  {542, L79}

\bibitem[\protect\citeauthoryear{{Gillmon} \& {Shull}}{{Gillmon} \&
  {Shull}}{2006}]{Gillmon2006}
{Gillmon} K.,  {Shull} J.~M.,  2006, \mn@doi [\apj] {10.1086/498055}, \href
  {http://adsabs.harvard.edu/abs/2006ApJ...636..908G} {636, 908}

\bibitem[\protect\citeauthoryear{{Gillmon}, {Shull}, {Tumlinson}  \&
  {Danforth}}{{Gillmon} et~al.}{2006}]{Gillmon2006b}
{Gillmon} K.,  {Shull} J.~M.,  {Tumlinson} J.,   {Danforth} C.,  2006, \mn@doi
  [\apj] {10.1086/498053}, \href
  {http://adsabs.harvard.edu/abs/2006ApJ...636..891G} {636, 891}

\bibitem[\protect\citeauthoryear{{Gordon}, {Witt}, {Carruthers}, {Christensen}
  \& {Dohne}}{{Gordon} et~al.}{1994}]{Gordon1994}
{Gordon} K.~D.,  {Witt} A.~N.,  {Carruthers} G.~R.,  {Christensen} S.~A.,
  {Dohne} B.~C.,  1994, \mn@doi [\apj] {10.1086/174602}, \href
  {http://cdsads.u-strasbg.fr/abs/1994ApJ...432..641G} {432, 641}

\bibitem[\protect\citeauthoryear{{Haikala}, {Mattila}, {Bowyer}, {Sasseen},
  {Lampton}  \& {Knude}}{{Haikala} et~al.}{1995}]{Haikala1995}
{Haikala} L.~K.,  {Mattila} K.,  {Bowyer} S.,  {Sasseen} T.~P.,  {Lampton} M.,
   {Knude} J.,  1995, \mn@doi [\apj] {10.1086/187829}, \href
  {http://cdsads.u-strasbg.fr/abs/1995ApJ...443L..33H} {443, L33}

\bibitem[\protect\citeauthoryear{{Hamden}, {Schiminovich}  \&
  {Seibert}}{{Hamden} et~al.}{2013}]{Hamden2013}
{Hamden} E.~T.,  {Schiminovich} D.,   {Seibert} M.,  2013, \mn@doi [\apj]
  {10.1088/0004-637X/779/2/180}, \href
  {http://adsabs.harvard.edu/abs/2013ApJ...779..180H} {779, 180}

\bibitem[\protect\citeauthoryear{{Henry}}{{Henry}}{1991}]{Henry1991}
{Henry} R.~C.,  1991, \mn@doi [\araa] {10.1146/annurev.aa.29.090191.000513},
  \href {http://adsabs.harvard.edu/abs/1991ARA%26A..29...89H} {29, 89}

\bibitem[\protect\citeauthoryear{{Henry} \& {Murthy}}{{Henry} \&
  {Murthy}}{1993}]{Henry1993}
{Henry} R.~C.,  {Murthy} J.,  1993, \mn@doi [\apjl] {10.1086/187105}, \href
  {http://adsabs.harvard.edu/abs/1993ApJ...418L..17H} {418, L17}

\bibitem[\protect\citeauthoryear{{Henry}, {Murthy}, {Overduin}  \&
  {Tyler}}{{Henry} et~al.}{2015}]{Henry2015}
{Henry} R.~C.,  {Murthy} J.,  {Overduin} J.,   {Tyler} J.,  2015, \mn@doi
  [\apj] {10.1088/0004-637X/798/1/14}, \href
  {http://adsabs.harvard.edu/abs/2015ApJ...798...14H} {798, 14}

\bibitem[\protect\citeauthoryear{{Henyey} \& {Greenstein}}{{Henyey} \&
  {Greenstein}}{1941}]{Henyey1941}
{Henyey} L.~G.,  {Greenstein} J.~L.,  1941, \mn@doi [\apj] {10.1086/144246},
  \href {http://cdsads.u-strasbg.fr/abs/1941ApJ....93...70H} {93, 70}

\bibitem[\protect\citeauthoryear{{Hurwitz}}{{Hurwitz}}{1998}]{Hurwitz1998}
{Hurwitz} M.,  1998, \mn@doi [\apjl] {10.1086/311387}, \href
  {http://adsabs.harvard.edu/abs/1998ApJ...500L..67H} {500, L67}

\bibitem[\protect\citeauthoryear{{Jo}, {Min}, {Lim}  \& {Seon}}{{Jo}
  et~al.}{2012}]{Jo2012}
{Jo} Y.-S.,  {Min} K.-W.,  {Lim} T.-H.,   {Seon} K.-I.,  2012, \mn@doi [\apj]
  {10.1088/0004-637X/756/1/38}, \href
  {https://ui.adsabs.harvard.edu/abs/2012ApJ...756...38J} {756, 38}

\bibitem[\protect\citeauthoryear{{Jo}, {Seon}, {Min}, {Edelstein}  \&
  {Han}}{{Jo} et~al.}{2017}]{Jo2017}
{Jo} Y.-S.,  {Seon} K.-I.,  {Min} K.-W.,  {Edelstein} J.,   {Han} W.,  2017,
  \mn@doi [\apjs] {10.3847/1538-4365/aa8091}, \href
  {http://adsabs.harvard.edu/abs/2017ApJS..231...21J} {231, 21}

\bibitem[\protect\citeauthoryear{{Jura}}{{Jura}}{1979}]{Jura1979}
{Jura} M.,  1979, \mn@doi [\apj] {10.1086/156788}, \href
  {http://adsabs.harvard.edu/abs/1979ApJ...227..798J} {227, 798}

\bibitem[\protect\citeauthoryear{{Lampton}, {Margon}  \& {Bowyer}}{{Lampton}
  et~al.}{1976}]{Lampton1976}
{Lampton} M.,  {Margon} B.,   {Bowyer} S.,  1976, \mn@doi [\apj]
  {10.1086/154592}, \href {http://cdsads.u-strasbg.fr/abs/1976ApJ...208..177L}
  {208, 177}

\bibitem[\protect\citeauthoryear{{Lee} et~al.,}{{Lee} et~al.}{2008}]{Lee2008}
{Lee} D.-H.,  et~al., 2008, \mn@doi [\apj] {10.1086/591778}, \href
  {http://cdsads.u-strasbg.fr/abs/2008ApJ...686.1155L} {686, 1155}

\bibitem[\protect\citeauthoryear{{Lim}, {Min}  \& {Seon}}{{Lim}
  et~al.}{2013}]{Lim2013}
{Lim} T.-H.,  {Min} K.-W.,   {Seon} K.-I.,  2013, \mn@doi [\apj]
  {10.1088/0004-637X/765/2/107}, \href
  {https://ui.adsabs.harvard.edu/abs/2013ApJ...765..107L} {765, 107}

\bibitem[\protect\citeauthoryear{{Low} et~al.,}{{Low} et~al.}{1984}]{Low1984}
{Low} F.~J.,  et~al., 1984, \mn@doi [\apjl] {10.1086/184213}, \href
  {http://cdsads.u-strasbg.fr/abs/1984ApJ...278L..19L} {278, L19}

\bibitem[\protect\citeauthoryear{{Madau}}{{Madau}}{1992}]{Madau1992}
{Madau} P.,  1992, \mn@doi [\apjl] {10.1086/186334}, \href
  {http://adsabs.harvard.edu/abs/1992ApJ...389L...1M} {389, L1}

\bibitem[\protect\citeauthoryear{{Markkanen}}{{Markkanen}}{1979}]{Markkanen1979}
{Markkanen} T.,  1979, \aap, \href
  {http://adsabs.harvard.edu/abs/1979A%26A....74..201M} {74, 201}

\bibitem[\protect\citeauthoryear{{Marshall}, {Robin}, {Reyl{\'e}}, {Schultheis}
   \& {Picaud}}{{Marshall} et~al.}{2006}]{Marshall2006}
{Marshall} D.~J.,  {Robin} A.~C.,  {Reyl{\'e}} C.,  {Schultheis} M.,   {Picaud}
  S.,  2006, \mn@doi [\aap] {10.1051/0004-6361:20053842}, \href
  {http://adsabs.harvard.edu/abs/2006A%26A...453..635M} {453, 635}

\bibitem[\protect\citeauthoryear{{Martin}, {Hurwitz}  \& {Bowyer}}{{Martin}
  et~al.}{1990}]{Martin1990}
{Martin} C.,  {Hurwitz} M.,   {Bowyer} S.,  1990, \mn@doi [\apj]
  {10.1086/168681}, \href {http://cdsads.u-strasbg.fr/abs/1990ApJ...354..220M}
  {354, 220}

\bibitem[\protect\citeauthoryear{{Martin}, {Hurwitz}  \& {Bowyer}}{{Martin}
  et~al.}{1991}]{Martin1991}
{Martin} C.,  {Hurwitz} M.,   {Bowyer} S.,  1991, \mn@doi [\apj]
  {10.1086/170527}, \href {http://adsabs.harvard.edu/abs/1991ApJ...379..549M}
  {379, 549}

\bibitem[\protect\citeauthoryear{{Martin} et~al.,}{{Martin}
  et~al.}{2005}]{Martin2005}
{Martin} D.~C.,  et~al., 2005, \mn@doi [\apjl] {10.1086/426387}, \href
  {http://cdsads.u-strasbg.fr/abs/2005ApJ...619L...1M} {619, L1}

\bibitem[\protect\citeauthoryear{{Mathis}, {Whitney}  \& {Wood}}{{Mathis}
  et~al.}{2002}]{Mathis2002}
{Mathis} J.~S.,  {Whitney} B.~A.,   {Wood} K.,  2002, \mn@doi [\apj]
  {10.1086/341007}, \href {http://adsabs.harvard.edu/abs/2002ApJ...574..812M}
  {574, 812}

\bibitem[\protect\citeauthoryear{{Mattila}}{{Mattila}}{1976}]{Mattila1976}
{Mattila} K.,  1976, \aap, \href
  {https://ui.adsabs.harvard.edu/abs/1976A&A....47...77M} {47, 77}

\bibitem[\protect\citeauthoryear{{Mattila} \& {V{\"a}is{\"a}nen}}{{Mattila} \&
  {V{\"a}is{\"a}nen}}{2019}]{Mattila2019}
{Mattila} K.,  {V{\"a}is{\"a}nen} P.,  2019, arXiv e-prints, \href
  {https://ui.adsabs.harvard.edu/abs/2019arXiv190508825M} {p. arXiv:1905.08825}

\bibitem[\protect\citeauthoryear{{Mattila}, {V{\"a}is{\"a}nen}, {Lehtinen},
  {von Appen-Schnur}  \& {Leinert}}{{Mattila} et~al.}{2017}]{Mattila2017}
{Mattila} K.,  {V{\"a}is{\"a}nen} P.,  {Lehtinen} K.,  {von Appen-Schnur} G.,
  {Leinert} C.,  2017, \mn@doi [\mnras] {10.1093/mnras/stx1296}, \href
  {https://ui.adsabs.harvard.edu/abs/2017MNRAS.470.2152M} {470, 2152}

\bibitem[\protect\citeauthoryear{{Mattila}, {Haas}, {Haikala}, {Jo},
  {Lehtinen}, {Leinert}  \& {V{\"a}is{\"a}nen}}{{Mattila}
  et~al.}{2018}]{Mattila2018}
{Mattila} K.,  {Haas} M.,  {Haikala} L.~K.,  {Jo} Y.~S.,  {Lehtinen} K.,
  {Leinert} C.,   {V{\"a}is{\"a}nen} P.,  2018, \mn@doi [\aap]
  {10.1051/0004-6361/201833196}, \href
  {https://ui.adsabs.harvard.edu/abs/2018A&A...617A..42M} {617, A42}

\bibitem[\protect\citeauthoryear{{Murthy}}{{Murthy}}{2014}]{Murthy2014a}
{Murthy} J.,  2014, \mn@doi [\apjs] {10.1088/0067-0049/213/2/32}, \href
  {http://adsabs.harvard.edu/abs/2014ApJS..213...32M} {213, 32}

\bibitem[\protect\citeauthoryear{{Murthy}}{{Murthy}}{2015}]{MurthyANSIC}
{Murthy} J.,  2015, {DiffuseModel: Modeling the diffuse ultraviolet
  background}, Astrophysics Source Code Library (\mn@eprint {ascl} {1512.012})

\bibitem[\protect\citeauthoryear{{Murthy}}{{Murthy}}{2016}]{Murthy2016}
{Murthy} J.,  2016, \mn@doi [\mnras] {10.1093/mnras/stw755}, \href
  {http://cdsads.u-strasbg.fr/abs/2016MNRAS.459.1710M} {459, 1710}

\bibitem[\protect\citeauthoryear{{Murthy}, {Im}, {Henry}  \&
  {Holberg}}{{Murthy} et~al.}{1993}]{Murthy1993b}
{Murthy} J.,  {Im} M.,  {Henry} R.~C.,   {Holberg} J.~B.,  1993, \mn@doi [\apj]
  {10.1086/173524}, \href
  {https://ui.adsabs.harvard.edu/abs/1993ApJ...419..739M} {419, 739}

\bibitem[\protect\citeauthoryear{{Neugebauer} et~al.,}{{Neugebauer}
  et~al.}{1984}]{Neugebauer1984}
{Neugebauer} G.,  et~al., 1984, \mn@doi [\apjl] {10.1086/184209}, \href
  {http://adsabs.harvard.edu/abs/1984ApJ...278L...1N} {278, L1}

\bibitem[\protect\citeauthoryear{{Onaka} \& {Kodaira}}{{Onaka} \&
  {Kodaira}}{1991}]{Onaka1991}
{Onaka} T.,  {Kodaira} K.,  1991, \mn@doi [\apj] {10.1086/170526}, \href
  {http://cdsads.u-strasbg.fr/abs/1991ApJ...379..532O} {379, 532}

\bibitem[\protect\citeauthoryear{{Perryman} et~al.,}{{Perryman}
  et~al.}{1997}]{Perryman1997}
{Perryman} M.~A.~C.,  et~al., 1997, \aap, \href
  {http://adsabs.harvard.edu/abs/1997A%26A...323L..49P} {323, L49}

\bibitem[\protect\citeauthoryear{{Planck Collaboration} et~al.,}{{Planck
  Collaboration} et~al.}{2011}]{Planck2011}
{Planck Collaboration} et~al., 2011, \mn@doi [\aap]
  {10.1051/0004-6361/201116485}, \href
  {https://ui.adsabs.harvard.edu/abs/2011A&A...536A..24P} {536, A24}

\bibitem[\protect\citeauthoryear{{Planck Collaboration} et~al.,}{{Planck
  Collaboration} et~al.}{2014}]{Planck2014}
{Planck Collaboration} et~al., 2014, \mn@doi [\aap]
  {10.1051/0004-6361/201323195}, \href
  {http://adsabs.harvard.edu/abs/2014A%26A...571A..11P} {571, A11}

\bibitem[\protect\citeauthoryear{{Puthiyaveettil}, {Murthy}  \&
  {Fix}}{{Puthiyaveettil} et~al.}{2010}]{Puthiyaveettil2010}
{Puthiyaveettil} S.,  {Murthy} J.,   {Fix} J.~D.,  2010, \mn@doi [\mnras]
  {10.1111/j.1365-2966.2010.17149.x}, \href
  {http://cdsads.u-strasbg.fr/abs/2010MNRAS.408...53P} {408, 53}

\bibitem[\protect\citeauthoryear{{Reach}, {Koo}  \& {Heiles}}{{Reach}
  et~al.}{1994}]{Reach1994}
{Reach} W.~T.,  {Koo} B.-C.,   {Heiles} C.,  1994, \mn@doi [\apj]
  {10.1086/174353}, \href {http://adsabs.harvard.edu/abs/1994ApJ...429..672R}
  {429, 672}

\bibitem[\protect\citeauthoryear{{Sasseen} \& {Deharveng}}{{Sasseen} \&
  {Deharveng}}{1996}]{Sasseen1996}
{Sasseen} T.~P.,  {Deharveng} J.~M.,  1996, \mn@doi [\apj] {10.1086/177815},
  \href {https://ui.adsabs.harvard.edu/abs/1996ApJ...469..691S} {469, 691}

\bibitem[\protect\citeauthoryear{{Savage}, {Bohlin}, {Drake}  \&
  {Budich}}{{Savage} et~al.}{1977}]{Savage1977}
{Savage} B.~D.,  {Bohlin} R.~C.,  {Drake} J.~F.,   {Budich} W.,  1977, \mn@doi
  [\apj] {10.1086/155471}, \href
  {http://adsabs.harvard.edu/abs/1977ApJ...216..291S} {216, 291}

\bibitem[\protect\citeauthoryear{{Schiminovich}, {Friedman}, {Martin}  \&
  {Morrissey}}{{Schiminovich} et~al.}{2001}]{Schiminovich2001}
{Schiminovich} D.,  {Friedman} P.~G.,  {Martin} C.,   {Morrissey} P.~F.,  2001,
  \mn@doi [\apjl] {10.1086/338656}, \href
  {http://adsabs.harvard.edu/abs/2001ApJ...563L.161S} {563, L161}

\bibitem[\protect\citeauthoryear{{Schlegel}, {Finkbeiner}  \&
  {Davis}}{{Schlegel} et~al.}{1998}]{Schlegel1998}
{Schlegel} D.~J.,  {Finkbeiner} D.~P.,   {Davis} M.,  1998, \mn@doi [\apj]
  {10.1086/305772}, \href {http://adsabs.harvard.edu/abs/1998ApJ...500..525S}
  {500, 525}

\bibitem[\protect\citeauthoryear{{Seon} et~al.,}{{Seon}
  et~al.}{2011}]{Seon2011}
{Seon} K.-I.,  et~al., 2011, \mn@doi [\apjs] {10.1088/0067-0049/196/2/15},
  \href {https://ui.adsabs.harvard.edu/abs/2011ApJS..196...15S} {196, 15}

\bibitem[\protect\citeauthoryear{{Shalima} \& {Murthy}}{{Shalima} \&
  {Murthy}}{2004}]{Shalima2004}
{Shalima} P.,  {Murthy} J.,  2004, \mn@doi [\mnras]
  {10.1111/j.1365-2966.2004.08022.x}, \href
  {https://ui.adsabs.harvard.edu/abs/2004MNRAS.352.1319S} {352, 1319}

\bibitem[\protect\citeauthoryear{{Shalima}, {Sujatha}, {Murthy}, {Henry}  \&
  {Sahnow}}{{Shalima} et~al.}{2006}]{Shalima2006}
{Shalima} P.,  {Sujatha} N.~V.,  {Murthy} J.,  {Henry} R.~C.,   {Sahnow} D.~J.,
   2006, \mn@doi [\mnras] {10.1111/j.1365-2966.2006.10071.x}, \href
  {https://ui.adsabs.harvard.edu/abs/2006MNRAS.367.1686S} {367, 1686}

\bibitem[\protect\citeauthoryear{{Sujatha}, {Shalima}, {Murthy}  \&
  {Henry}}{{Sujatha} et~al.}{2005}]{Sujatha2005}
{Sujatha} N.~V.,  {Shalima} P.,  {Murthy} J.,   {Henry} R.~C.,  2005, \mn@doi
  [\apj] {10.1086/444532}, \href
  {https://ui.adsabs.harvard.edu/abs/2005ApJ...633..257S} {633, 257}

\bibitem[\protect\citeauthoryear{{Sujatha}, {Murthy}, {Karnataki}, {Henry}  \&
  {Bianchi}}{{Sujatha} et~al.}{2009}]{Sujatha2009}
{Sujatha} N.~V.,  {Murthy} J.,  {Karnataki} A.,  {Henry} R.~C.,   {Bianchi} L.,
   2009, \mn@doi [\apj] {10.1088/0004-637X/692/2/1333}, \href
  {https://ui.adsabs.harvard.edu/abs/2009ApJ...692.1333S} {692, 1333}

\bibitem[\protect\citeauthoryear{{Sujatha}, {Murthy}, {Suresh}, {Conn Henry}
  \& {Bianchi}}{{Sujatha} et~al.}{2010}]{Sujatha2010}
{Sujatha} N.~V.,  {Murthy} J.,  {Suresh} R.,  {Conn Henry} R.,   {Bianchi} L.,
  2010, \mn@doi [\apj] {10.1088/0004-637X/723/2/1549}, \href
  {http://cdsads.u-strasbg.fr/abs/2010ApJ...723.1549S} {723, 1549}

\bibitem[\protect\citeauthoryear{{Tennyson}, {Henry}, {Feldman}  \&
  {Hartig}}{{Tennyson} et~al.}{1988}]{Tennyson1988}
{Tennyson} P.~D.,  {Henry} R.~C.,  {Feldman} P.~D.,   {Hartig} G.~F.,  1988,
  \mn@doi [\apj] {10.1086/166481}, \href
  {http://cdsads.u-strasbg.fr/abs/1988ApJ...330..435T} {330, 435}

\bibitem[\protect\citeauthoryear{{Voyer}, {Gardner}, {Teplitz}, {Siana}  \& {de
  Mello}}{{Voyer} et~al.}{2011}]{Voyer2011}
{Voyer} E.~N.,  {Gardner} J.~P.,  {Teplitz} H.~I.,  {Siana} B.~D.,   {de Mello}
  D.~F.,  2011, \mn@doi [\apj] {10.1088/0004-637X/736/2/80}, \href
  {http://adsabs.harvard.edu/abs/2011ApJ...736...80V} {736, 80}

\bibitem[\protect\citeauthoryear{{Weiland}, {Blitz}, {Dwek}, {Hauser},
  {Magnani}  \& {Rickard}}{{Weiland} et~al.}{1986}]{Weiland1986}
{Weiland} J.~L.,  {Blitz} L.,  {Dwek} E.,  {Hauser} M.~G.,  {Magnani} L.,
  {Rickard} L.~J.,  1986, \mn@doi [\apj] {10.1086/184714}, \href
  {https://ui.adsabs.harvard.edu/\#abs/1986ApJ...306L.101W} {306, L101}

\bibitem[\protect\citeauthoryear{{Witt} \& {Petersohn}}{{Witt} \&
  {Petersohn}}{1994}]{Witt1994}
{Witt} A.~N.,  {Petersohn} J.~K.,  1994, in {Cutri} R.~M.,  {Latter} W.~B.,
  eds,  Astronomical Society of the Pacific Conference Series Vol. 58, The
  First Symposium on the Infrared Cirrus and Diffuse Interstellar Clouds. p.~91

\bibitem[\protect\citeauthoryear{{Witt}, {Petersohn}, {Holberg}, {Murthy},
  {Dring}  \& {Henry}}{{Witt} et~al.}{1993}]{Witt1993}
{Witt} A.~N.,  {Petersohn} J.~K.,  {Holberg} J.~B.,  {Murthy} J.,  {Dring} A.,
   {Henry} R.~C.,  1993, \mn@doi [\apj] {10.1086/172788}, \href
  {http://adsabs.harvard.edu/abs/1993ApJ...410..714W} {410, 714}

\bibitem[\protect\citeauthoryear{{Witt}, {Friedmann}  \& {Sasseen}}{{Witt}
  et~al.}{1997}]{Witt1997}
{Witt} A.~N.,  {Friedmann} B.~C.,   {Sasseen} T.~P.,  1997, \mn@doi [\apj]
  {10.1086/304093}, \href {http://cdsads.u-strasbg.fr/abs/1997ApJ...481..809W}
  {481, 809}

\bibitem[\protect\citeauthoryear{{Xu} et~al.,}{{Xu} et~al.}{2005}]{Xu2005}
{Xu} C.~K.,  et~al., 2005, \mn@doi [\apjl] {10.1086/425252}, \href
  {http://adsabs.harvard.edu/abs/2005ApJ...619L..11X} {619, L11}

\makeatother
\end{thebibliography}

\bsp	
\label{lastpage}
\end{document}